\documentclass[twoside,reqno,11pt]{article}
\usepackage{amsmath,amssymb,euscript,bbold}
\usepackage{array}
\usepackage[T1]{fontenc}

\DeclareMathOperator{\ads}{adS}
\DeclareMathOperator{\vol}{dvol}
\newcommand{\1}{\mathbb{1}}
\newcommand{\CC}{\mathbb{C}}
\newcommand{\CP}{\mathbb{CP}}
\newcommand{\Cl}{\mathrm{C}\ell}
\newcommand{\D}{\mathsf{D}}
\newcommand{\EE}{\mathbb{E}}
\newcommand{\GL}{\mathrm{GL}}
\newcommand{\M}{\mathsf{M}}
\newcommand{\NN}{\mathbb{N}}
\newcommand{\RP}{\mathbb{RP}}
\newcommand{\RR}{\mathbb{R}}
\newcommand{\R}{\mathsf{R}}
\newcommand{\SO}{\mathrm{SO}}
\newcommand{\SU}{\mathrm{SU}}
\newcommand{\Spin}{\mathrm{Spin}}
\newcommand{\TT}{\mathbb{T}}
\newcommand{\U}{\mathrm{U}}
\newcommand{\ZZ}{\mathbb{Z}}
\newcommand{\bx}{\boldsymbol{x}}
\newcommand{\eF}{\EuScript{F}}

\newcommand{\eN}{\EuScript{N}}
\newcommand{\fg}{\mathfrak{g}}
\newcommand{\half}{\tfrac{1}{2}}
\newcommand{\osp}{\mathfrak{osp}}
\newcommand{\real}[1]{[\![#1]\!]}
\newcommand{\repre}[1]{\underline{\boldsymbol{#1}}}
\newcommand{\so}{\mathfrak{so}}
\newcommand{\su}{\mathfrak{su}}
\renewcommand{\O}{\mathrm{O}}
\renewcommand{\Sp}{\mathrm{Sp}} 
\renewcommand{\d}{\partial}
\renewcommand{\sp}{\mathfrak{sp}}
\renewcommand{\u}{\mathfrak{u}}
\begin{document}


\thispagestyle{empty}
\leftline{\copyright~ 1998 International Press}
\leftline{Adv. Theor. Math. Phys. {\bf 2} (1998) 1249-1286}  
\vspace{0.4in}
\centerline {\huge \bf Branes at conical singularities}
\medskip
\centerline {\huge \bf and holography} 
\begin{center}
{\bf B.S. Acharya, J. M. Figueroa-O'Farrill, C.M. Hull, B. Spence}
\vspace{0.2in}
 
Department of Physics \\
Queen Mary and Wesfield College\\
Mile End Road\\
London E1 4NS, UK\\
{\tt{r.acharya,j.m.figueroa,c.m.hull,b.spence}@qmw.ac.uk}

\renewcommand{\thefootnote}{}
\footnotetext{e-print archive: {\texttt 
http://xxx.lanl.gov/abs/hep-th/9705006}}
\renewcommand{\thefootnote}{\arabic{footnote}}

\vskip0.4in
{\bf Abstract}
\vskip0.1in
\parbox[c]{4.5in}{\small \hspace{.2in} For supergravity solutions which are the product of an anti-de Sitter
space with an Einstein space $X$, we study the relation between the
amount of supersymmetry preserved and the geometry of $X$.  Depending
on the dimension and the amount of supersymmetry, the following
geometries for $X$ are possible, in addition to the maximally
supersymmetric spherical geometry: Einstein-Sasaki in dimension
$2k{+}1$, $3$-Sasaki in dimension $4k{+}3$, $7$-dimensional manifolds
of weak $G_2$ holonomy and $6$-dimensional nearly K\"ahler manifolds.
Many new examples of such manifolds are presented which are not
homogeneous and hence have escaped earlier classification efforts.
String or $\M$ theory in these vacua are conjectured to be dual to
superconformal field theories.  The brane solutions interpolating
between these anti-de Sitter near-horizon geometries and the product
of Minkowski space with a cone over $X$ lead to an interpretation of
the dual superconformal field theory as the world-volume theory for
branes at a conical singularity (cone branes).  We propose a
description of those field theories whose associated cones are
obtained by (hyper-)K\"ahler quotients.}
\end{center}

\newpage

\pagestyle{myheadings}
\markboth{\it BRANES AT CONICAL SINGULARITIES AND HOLOGRAPHY}{\it B. S. ACHARYA, J. M. FIGUEROA, C.M. HULL, B. SPENCE }
\section{Introduction and Motivation}
Maldacena \cite{Malda} has conjectured that the 't~Hooft large $N$
limit of $\eN{=}4$ supersymmetric Yang--Mills with gauge group $\SU_N$
is dual to type IIB superstring theory on $\left(\ads_5 \times
S^5\right)_N$, where the subscript indicates that the sizes of the
spaces grow with $N$.  A more precise version of the conjecture was
formulated in \cite{GKP,WittenAdS} where a simple recipe was given for
relating gauge theory correlators to string theory S-matrix elements,
and these are given in terms of classical supergravity in the large
$N$ limit.

\pagenumbering{arabic}
\setcounter{page}{1250}
This conjecture was motivated by considering $N$ parallel $\D3$-branes
for $N$ large and taking a limit in which the gauge theory on the
brane decouples from the physics of the bulk \cite{Malda}.  When
$g_sN$ is small (with $g_s$ the string coupling), the system is
well-described by $4$-dimensional super Yang-Mills theory with $\SU_N$
gauge group, while if $g_sN$ is large, the system is described by IIB
string theory in the near-horizon geometry, which is $\ads_5
\times S^5$.  These are then dual descriptions of the same system,
leading to the conjectured equivalence.  String loop corrections
correspond to $\frac1{N}$ corrections in the gauge theory so that in
the large $N$ limit, one can use classical supergravity theory in
the $\ads_5 \times S^5$ background.

This can be generalised to any $p$-brane of superstring theory or $\M$
theory and this leads to a relation between the worldvolume theory
with $\SU_N$ gauge symmetry and the string or $\M$ theory in the
space-time which arises in the near-horizon limit of the $p$-brane
spacetime \cite{IMSY}.  Of particular interest are those cases---the
$\D3$-brane and the $\M2$ and $\M5$-branes---in which the near-horizon
geometry is a supersymmetric anti-de~Sitter space solution of the form
$\ads_{p+2} \times S^d$ ($d\equiv D-p-2$) and the worldvolume theory
is a superconformal field theory.  In these cases, there is a
holographic duality between the string or $\M$ theory in
anti-de~Sitter space and the superconformal field theory (which may be
thought of as being at the boundary of the anti-de~Sitter space
\cite{Malda,WittenAdS}).  The superconformal symmetry in $p+1$
dimensions is identified with the anti-de~Sitter supersymmetry in
$p+2$ dimensions.

An interesting extension of these ideas is to explore $p$-branes which
exhibit near-horizon geometries of the form $\ads_{p+2} \times X_d$,
where $X$ is an Einstein manifold.  If the geometry is supersymmetric,
then the string or $\M$ theory in that background is expected to be
holographically dual to a superconformal field theory.  These vacua
are not maximally supersymmetric unless $X$ is a round sphere or the
real projective space, which is the near-horizon geometry of branes on
an orientifold \cite{WittenBaryons}, and as a result the dual field
theories have less than maximal supersymmetry.  Many such cases have
been studied already, particularly in the context of type IIB
superstring theory, but also in $\M$ theory \cite{AOY-MCFT}.  A simple
modification is to let $X = S^d/\Gamma$ be a finite quotient of the
sphere \cite{KaSi-orbifolds, LNV, OT-orbi, Berkooz-S4orbi, AOT-orbi,
Gomis1, Gomis2}.  Alternatively, if we recall that $S^{2n+1}$ is a
circle bundle over $\CP^n$, one can replace $\CP^n$ with other
K\"ahler--Einstein $n$-folds.  This was done in the type IIB context
(i.e., $n=2$) in \cite{Kehagias}, generalising \cite{PopeWarner}.
Another generalisation is to replace $S^d \cong \SO_{d+1}/\SO_d$ with
another homogeneous space $G/H$. Homogeneous vacua of supergravity
theories were studied intensively in the early days of Kaluza--Klein
supergravity (see, e.g., \cite{DNP}).  There is a complete
classification in dimension seven \cite{CastellaniRomansWarner} and a
partial list in dimension five \cite{Romans5}.  One such
five-dimensional example is $T^{1,1} \equiv (\SU_2 \times
\SU_2)/\U_1$, whose dual conformal field theory was recently discussed
by Klebanov and Witten \cite{KlebanovWitten} (see also
\cite{Gubser-Einstein}).  They interpreted the vacuum in terms of the
near-horizon geometry of parallel $\D3$-branes at a conical
singularity of a Calabi--Yau threefold.  Similar ideas have been
considered in \cite{Morrison}.

The purpose of the present paper, which subsumes \cite{JMFSUSY98}, is
to study those $\M$ theory or superstring vacua of the form
$\ads_{p+2}\times X_d$ that preserve some supersymmetry.  Our aim will
be to understand the constraints supersymmetry imposes on the geometry
of the Einstein manifolds $X_d$, and then to use the geometry of these
manifolds in order to identify the superconformal symmetries which are
required under the adS/CFT correspondence.  This is certainly likely
to be useful in understanding {\em generic\/} features of this
correspondence in supersymmetric cases.

Any solution of the form $\ads_{p+2}\times X_d$ is the near-horizon
geometry of a $p$-brane solution \cite{DLPS} which will be
supersymmetric if the near-horizon geometry is, and which interpolates
between $\ads_{p+2}\times X_d$ and a vacuum $M_{p+1}\times C(X)_{d+1}$
where $C(X)_{d+1}$ is a cone over $X$ (defined below) and $M_{p+1}$ is
($p+1$)-dimensional Minkowski space. The case in which $X_d$ (with
$d{=}7$) is a coset space was considered in \cite{CCDAFFT}, but we
will consider general Einstein spaces $X_d$ which in addition preserve
some supersymmetries.  The cone $C(X)$ is Ricci-flat and the number of
supersymmetries of the $M_{p+1}\times C(X)_{d+1}$ vacuum depends on
the number of covariantly constant spinors. This is determined by the
holonomy group of $C(X)$, and such holonomies have been classified
(see, e.g., \cite{Wang}).  The number of supersymmetries on
$\ads_{p+2}\times X_d$ depends on the number of Killing spinors on
$X$, but these all arise from covariantly constant spinors on $C(X)$
\cite{Baer}, leading to a characterisation of supersymmetric
anti-de~Sitter solutions.

We will show that for those vacua $\ads_{p+2}\times X_d$ which are
supersymmetric, the geometry of $X$ is highly constrained.  Depending
on $d\equiv D{-}p{-}2$, the possible geometries of $X$ are as follows:
nearly K\"ahler for $d=6$, weak $G_2$ holonomy for $d{=}7$,
Einstein--Sasaki for $d{=}2k+1$, and $3$-Sasaki for $d{=}4k+3$; this
gives many supersymmetric compactifications of string and $\M$ theory
which have not been considered previously.

Given a $p$-brane solution of the above type, the interpolating
solution argument \cite{Malda} then leads to a conjectured duality
between the string or $\M$ theory in a supersymmetric background of
the form $\ads_{p+2}\times X_d$ and a $(p+1)$-dimensional
superconformal field theory on the worldvolume of $N$ coincident
$p$-branes located at the conical singularity of the $M_{p+1}\times
C(X)_{d+1}$ vacuum.  The detailed description of the dual theories
will be left to future investigations, and in this paper we will focus
instead on generic features of the field theories which follow from
the common properties of each of the geometries listed above.

This paper is organised as follows.  In Section 2 we set the notation
concerning the supersymmetric branes which we will study in this paper
and we review the simple solutions with spherical near-horizon
geometries.  In Section 3 we discuss the solutions which can be
interpreted as branes sitting at a conical singularity in a Ricci-flat
manifold $C$, and we characterise the supersymmetric solutions in
terms of the holonomy of $C$.  In Section 4 we discuss their
near-horizon geometries in detail.  Section 5 contains many examples
including those in the early Kaluza--Klein literature as well as some
more recent ones which have hereto not been considered in relation
with supergravity.  As an application of our results, in Section 6 we
describe how the near-horizon geometry induces the superconformal
symmetry of the dual theory.  Finally in Section 7 we offer some
conclusions.

\section{Supersymmetric Branes and their Near-Horizon Geometries}

In this section we review the near-horizon geometries of the
elementary brane solutions of string and $\M$ theory.  This section
serves mostly to establish the notation.

\subsection{$\M$-branes}

Eleven-dimensional supergravity consists of the following fields
\cite{Nahm,CJS}: a Lorentzian metric $g$, a closed $4$-form $F$ with
a quantised flux and a gravitino $\Psi$.  By a supersymmetric vacuum
we will mean any solution of the equations of motion with $\Psi=0$ for
which the supersymmetry variation $\delta_\varepsilon \Psi=0$,
regarded as a linear equation on the spinor parameter $\varepsilon$,
has nontrivial solutions.  The eleven-dimensional spinorial
representation is $32$-dimensional and real, so there at most $32$
linearly independent solutions.  An important physical invariant of a
supersymmetric vacuum is the fraction $\nu \equiv \frac{1}{32}n$ of
the supersymmetry that the solution preserves.  For example, $F=0$ and
$g$ the flat metric on eleven-dimensional Minkowski spacetime is a
supersymmetric vacuum with $\nu=1$; that is, it is maximally
supersymmetric.  Other maximally supersymmetric vacua are
$\ads_4\times S^7$ and $\ads_7 \times S^4$ with $\star F$ and $F$
having quantised flux on the $S^7$ and $S^4$, respectively.

Eleven-dimensional supergravity has four types of elementary solutions
with $\nu=\half$: the $\M$-wave \cite{Mwave} and the Kaluza--Klein
monopole \cite{KKmonopoleGP,KKmonopoleS,Gbranes}, and the elementary
brane solutions: the $\M2$-brane \cite{DS2brane} and the $\M5$-brane
\cite{Guven}.  There should also be an $\M9$-brane solution
\cite{Gbranes,M9brane}.  In what follows we will focus on the $\M2$-
and $\M5$-branes.

\subsubsection{The $\M2$-brane}

The following background describes a number of parallel $\M2$-branes
\cite{DS2brane}:
\begin{align}\label{eq:M2}
g &= H^{-\frac23}\, g_{2+1} + H^{\frac13} g_8\\
F &= \pm \vol_{2+1} \wedge dH^{-1}\notag~,
\end{align}
where $g_{2+1}$ and $\vol_{2+1}$ are the metric and volume form on the
three-dimen-sional Minkowskian worldvolume $\EE^{2,1}$ of the branes;
$g_8$ is the metric on the eight-dimensional euclidean space $\EE^8$
transverse to the branes; and $H$ is a harmonic function on
$\EE^8$. For example, if we demand that $H$ depend only on the
transverse radial coordinate $r$ on $\EE^8$, we then find that
\begin{equation}\label{eq:H(r)}
H(r) = 1 + \frac{a^6}{r^6}~; \qquad  a^6 \equiv 2^5\pi^2 N \ell_p^6
\end{equation}
is the only solution with $\lim_{r\to\infty} H(r) = 1$.  This
corresponds to $N$ coincident membranes at $r=0$.  Here $\ell_p$ is
the eleven-dimensional Planck length.

Other choices of $H$ are possible.  For example one can consider
multicentre generalisations, where $H(\bx)$ is an arbitrary harmonic
function on $\EE^8$ with pointlike singularities and suitable
asymptotic behaviour $H(\bx) \to 1$, say, as $|\bx|\to\infty$.  This
corresponds to parallel branes localised at the singularities of $H$.
More generally, we can take $H$ to be invariant under some subgroup
$G$ of isometries of $g_8$.  These solutions correspond to branes
which are `delocalised' or smeared on the $G$-orbits.

We can easily determine the fraction $\nu$ of the supersymmetry which
the above solution preserves \cite{DS2brane}.  The supersymmetry
variation of the gravitino in a bosonic background $(g,F)$ is given by
\begin{equation}\label{eq:gravitinoshift}
\delta_\varepsilon\Psi_M = \nabla_M \varepsilon - \tfrac{1}{288}
\left( \Gamma_M{}^{PQRS} - 8 \delta_M{}^P\Gamma^{QRS}\right)
F_{PQRS}\,\varepsilon~,
\end{equation}
where $\nabla_M = \d_M + \tfrac14 \omega_M{}^{PQ}\Gamma_{PQ}$ is the
spin connection.  In the $\M2$-brane Ansatz given by \eqref{eq:M2} and
\eqref{eq:H(r)}, equation \eqref{eq:gravitinoshift} is solved by
\begin{equation*}
\varepsilon = H^{-\frac16}\, \varepsilon_\infty~,
\end{equation*}
where, choosing $x^0,x^1,x^2$ to be the directions tangent to the
worldvolume of the brane, $\varepsilon_\infty$ obeys
\begin{equation*}
\Gamma_{012}\,\varepsilon_\infty = \varepsilon_\infty~.
\end{equation*}
Because $\Gamma_{012}$ squares to the identity and is traceless, we
see that $\nu=\half$.  Nevertheless, the $\M2$-brane interpolates
between two maximally supersymmetric solutions: flat Minkowski space
$\EE^{10,1}$ infinitely far away from the brane, and $\ads_4 \times
S^7$ near the brane horizon
\cite{GibbonsTownsend,DuffGibbonsTownsend}.

Indeed, notice that the metric on the transverse space is given by
\begin{equation}\label{eq:EScone}
g_8 = dr^2 + r^2\,g_S~,
\end{equation}
where $g_S$ is the metric on the unit sphere $S^7 \subset \EE^8$.  In
the near-horizon region $r \ll a$,
\begin{equation*}
H(r) \sim \frac{a^6}{r^6}~,
\end{equation*}
so that the membrane metric becomes
\begin{equation*}
g = a^{-4}r^4 g_{2+1} + a^2 r^{-2} dr^2 + a^2 g_S~.
\end{equation*}
The last term is the metric on a round $S^7$ of radius $a=(2^5\pi^2
N)^{\frac16}\ell_p$; whereas the first two combine to produce the
metric on $4$-dimensional anti-de~Sitter spacetime with ``radius''
$R_{\ads} = \half a$:
\begin{equation}\label{eq:ads4BR}
g_{\ads} = R^2_{\ads} \left[ \frac{du^2}{u^2} + \left(
\frac{u}{R_{\ads}} \right)^2 \frac{g_{2+1}}{R_{\ads}^2} \right]~,
\end{equation}
with $u = \frac14 r^2/R_{\ads}$.

\subsubsection{The $\M5$-brane}

The $\M5$-brane is the magnetic dual to the $\M2$-brane.  The
background describing a number of parallel $\M5$-branes is given by
\cite{Guven}:
\begin{align}\label{eq:M5}
g &= H^{-\frac13}\, g_{5+1} + H^{\frac23} g_5\\
F &= \pm 3 \star_5 dH\notag~,
\end{align}
where $g_{5+1}$ and $\vol_{5+1}$ are the metric and volume form on the
six-dimensional Minkowskian worldvolume $\EE^{5,1}$ of the branes;
$g_5$ and $\star_5$ are the metric and Hodge operator on the
five-dimensional euclidean space $\EE^5$ transverse to the branes; and
$H$ is a harmonic function on $\EE^5$.  For example, if we demand that
$H$ depend only on the transverse radial coordinate $r$ on $\EE^5$, we
then find that
\begin{equation}\label{eq:H(r)5}
H(r) = 1 + \frac{a^3}{r^3}~; \qquad a^3 \equiv \pi N \ell_p^3
\end{equation}
is the only solution with $\lim_{r\to\infty} H(r) = 1$.  This
corresponds to $N$ coincident fivebranes at $r=0$.

As for the membrane solution, multicentre and `delocalised' fivebrane
solutions also exist.

The fraction $\nu$ of the supersymmetry which is preserved can be
computed as before \cite{Guven}.  The gravitino shift equation in the
bosonic background given by $(g,F)$ above is solved by
\begin{equation*}
\varepsilon = H^{-\frac1{12}}\, \varepsilon_\infty~,
\end{equation*}
where, choosing $x^0,x^1,\ldots,x^5$ to be the directions tangent to
the worldvolume of the fivebrane, $\varepsilon_\infty$ obeys
\begin{equation*}
\Gamma_{012345}\,\varepsilon_\infty = \varepsilon_\infty~.
\end{equation*}
$\Gamma_{012345}$ squares to the identity and is traceless, so that
again $\nu=\half$.  Nevertheless, as for the $\M2$-brane, the
$\M5$-brane also interpolates between two maximally supersymmetric
solutions: flat Minkowski space $\EE^{10,1}$ infinitely far away from
the brane, and $\ads_7 \times S^4$ near the brane horizon
\cite{GibbonsTownsend,DuffGibbonsTownsend}.

Indeed,
\begin{equation}\label{eq:ES3cone}
g_5 = dr^2 + r^2\,g_S~,
\end{equation}
where $g_S$ is now the metric on the unit sphere $S^4 \subset \EE^5$.
In the near-horizon region $r \ll a$,
\begin{equation*}
H(r) \sim \frac{a^3}{r^3}~,
\end{equation*}
so that the fivebrane metric becomes
\begin{equation*}
g \sim a^{-1} r g_{5+1} +a^2 r^{-2} dr^2 + a^2
g_S~.
\end{equation*}
The last term is the metric on a round $S^4$ of radius $a=(\pi
N)^{\frac13}\ell_p$, whereas the first two combine to produce the
metric on $7$-dimensional anti-de~Sitter spacetime with ``radius''
$R_{\ads} = 2 a$, analogous to \eqref{eq:ads4BR},
\begin{equation*}
g_{\ads} = R^2_{\ads} \left[ \frac{du^2}{u^2} + \left(
\frac{u}{R_{\ads}} \right)^2 \frac{g_{5+1}}{R_{\ads}^2} \right]~,
\end{equation*}
with now $u^2 = 2 R_{\ads}\,r$.

\subsection{String Branes}

The near-horizon geometries of $p$-branes in type II string theory are
not of the form
\begin{equation}\label{eq:adsxS}
\ads_{p+2} \times S^{D-p-2}~,
\end{equation}
with the exception of the $\D3$-brane of type IIB string theory,  
for which the near-horizon geometry
is $\ads _5 \times S^5$. For
other $\D{p}$-branes ($p\neq3$) the near-horizon geometry is conformal
to \eqref{eq:adsxS}, the conformal factor being nontrivial, and
either singular for $p<3$ or zero for $p>3$ as $r\to 0$.

Geometries of the form
\begin{equation}\label{eq:adsxSasas}
\ads_{p+2} \times S^{d} \times M_{D-p-2}
\end{equation}
for some space $M$ also arise, and compactifying on $M$ leads to a
$\ads_{p+2} \times S^{d}$ geometry. For example, $\ads_{3} \times
S^{3}$ arises from a D1-brane lying inside a D5-brane \cite{Malda},
while $\ads_{2} \times S^{2}$ is the near-horizon geometry for the
extreme Reissner-Nordstr\"om black hole.

\subsubsection{The $\D3$-brane in Type IIB}

The metric for the $\D3$-brane of type IIB is given by
\cite{HorowitzStrominger},
\begin{equation}\label{eq:D3}
g = H^{-\half}\, g_{3+1} + H^{\half}\, g_6~,
\end{equation}
where $g_{3+1}$ is the metric on the Minkowski worldvolume of the
brane $\EE^{3,1}$ and $g_6$ is the euclidean metric on the transverse
$\EE^6$.  The self-dual $5$-form $F$ has (quantised) flux on the unit
transverse five-sphere $S^5\subset \EE^6$ and the dilaton is constant.
$H$ is again harmonic in $\EE^6$ and the unique spherically symmetric
solution with $\lim_{r\to\infty} H(r) =1$ is
\begin{equation*}
H(r) = 1 + \frac{a^4}{r^4}~; \qquad a^4 \equiv 4\pi g N \ell_s^4
\end{equation*}
where $g$ is the  string coupling constant, given by the
exponential of the constant dilaton, and $\ell_s = \sqrt{\alpha'}$ is
the string length.  The solution corresponds to $N$ parallel
$\D3$-branes at $r=0$.  The ten-dimensional Planck length is $\ell_p
\equiv g^{\tfrac14} \ell_s$, so that $a$ can be rewritten as
\begin{equation}\label{eq:H(r)D3}
a^4 =  4\pi N \ell_p^4~.
\end{equation}

The near-horizon analysis is similar to that for the $\M2$ and $\M5$
branes above, and the near-horizon geometry is
\begin{equation*}
\ads_5 \times S^5~,
\end{equation*}
where the anti-de~Sitter and sphere radii are now equal, $R_{\ads} = a 
= \left(4\pi N\right)^{\frac14} \ell_p$.

\section{Branes at Conical Singularities}

All the branes considered in the previous section interpolate between
flat Minkowski spacetime asymptotically far away from the brane and
$\ads_{p+2} \times S^d$ near the brane horizon.  Both of these vacua
are maximally supersymmetric and the branes themselves preserve
$\nu=\half$ of the supersymmetry.  In this section we discuss the
brane solutions of \cite{DLPS} that interpolate between near-horizon
geometries of the form $\ads_{p+2} \times X $ and an asymptotic
geometry of the form $M_{p+1}\times C(X)$, where $X$ is a
$d$-dimensional Einstein manifold and $C(X)$ is the cone over $X$.
This can be interpreted as a set of coincident branes at a conical
singularity in a Ricci-flat manifold.  These brane solutions
interpolate between vacua which are not maximally supersymmetric and
the brane solutions themselves will preserve a smaller fraction
$\nu<\half$ of the supersymmetry.  In this section we will
characterise those branes with $\nu\neq0$ in terms of the holonomy
group of the transverse space.

\subsection{Cone Branes}

The $p$-brane solutions above all have metrics of the form
\begin{equation}\label{eq:pbranemetric}
g = H^{-\alpha}\,g_{p+1} + H^{\frac2\beta} \,g_E~,
\end{equation}
where
\begin{equation}\label{eq:H(r)ig}
H(r) = 1 + \left(\frac{a}{r}\right)^\beta~,
\end{equation}
$g_{p+1}$ is the Minkowski metric on the worldvolume of the $p$-brane,
and where $g_E$ is the flat Euclidean metric on $\EE^{D-p-1}$.  For
the $\M2$-brane, $\M5$-brane, and $\D3$-brane, $\beta =6,3,4$
respectively, and $\alpha =1 - \frac2\beta$.  The Euclidean metric
$g_E$ can be written as
\begin{equation}\label{eq:metricone}
g_E = dr^2 + r^2 g_S
\end{equation}
where $g_S$ is the round metric on the unit sphere $S^d \subset
\EE^d$ and $r$ is a radial coordinate. In the near-horizon limit in
which the constant term in $H$ can be dropped, the metric
\eqref{eq:pbranemetric} becomes
\begin{equation*}
g \sim \left(\frac{r}{a}\right)^{\alpha\beta} \,g_{p+1} +a^2
r^{-2} dr^2 + a^2 \,g_S~,
\end{equation*}
which is the metric on $\ads_{p+2} \times S^d$,
as can be seen by changing variables to $u = R_{\ads}
\left(\frac{r}{a}\right)^{\frac{\alpha\beta}{2}}$, where the anti-de
Sitter radius is $R_{\ads}=\frac{2a}{\alpha\beta}$.

Replacing the sphere $S^d$ by any other $d$-dimensional Einstein
manifold $X_d$ with the same cosmological constant $\Lambda = d-1$
gives another solution of the field equations on $\ads_{p+2} \times
X_d$, and we will be interested in those choices of $X$ that give
spontaneous compactifications to anti-de~Sitter space that preserve
some nonzero fraction $\nu$ of the supersymmetry.  Note that as $X$ is
a complete Einstein space with positive cosmological constant, it
follows by Myer's Theorem that it is compact (see, e.g.,
\cite{CheegerEbin}). There is a brane solution of the form \cite{DLPS}
\begin{equation}\label{eq:interpolatingbrane}
g = H^{-\alpha}\,g_{p+1} + H^{\frac2\beta}\,g_C~,
\end{equation}
where $H(r)$ is again given by \eqref{eq:H(r)ig} and the metric $g_C$
is given by replacing the spherical metric $g_S$ with the metric of
$X_d$ in \eqref{eq:metricone} to give
\begin{equation*}
g_C = dr^2 + r^2 g_X~.
\end{equation*}
The transverse space to the brane is now the metric cone $C \equiv
C(X)$ over $X$ with topology $C \cong \RR^+ \times X$, with $\RR^+$
the open half-line $0<r<\infty$ and metric $g_C$.  Since $X$ is
Einstein with $\Lambda = \dim X -1$, it follows that $C$ is
Ricci-flat; however unlike the case of the sphere, where $C$ is
actually flat, $C$ is now metrically singular at the apex of the cone
$r=0$.

In the near-horizon region $r\ll a$, the constant term in $H$ can be
dropped and the metric becomes approximately
\begin{equation*}
g \sim \left(\frac{a}{r}\right)^{\alpha\beta} \,g_{p+1} +a^2 r^{-2}
dr^2 + a^2 \,g_C~,
\end{equation*}
which is the metric on $\ads_{p+2} \times X_d$.  For large $r$, $H
\sim 1$ and the metric becomes
\begin{equation*}
g \sim g_{p+1}+g_C~,
\end{equation*}
on the product $M_{p+1}\times C(X)$ of ($p+1$)-dimensional Minkowski
space with the cone $C(X)$.  We interpret these solutions as
describing coincident branes located at the conical singularity of
$C(X)$, or {\em cone branes\/} for short. Note that whereas the
solution $M_{p+1}\times C(X)$ has a conical singularity at $r=0$, the
brane metric \eqref{eq:interpolatingbrane} approaches the metric of
$\ads_{p+2} \times X_d$ as $r \to 0$, which is non-singular at $r=0$.
However $r=0$ is a horizon for the brane and the solution can be
continued through the horizon.  In general there will be a singularity
inside the horizon.

These solutions can in principle be generalised by replacing $H$ with
more general harmonic functions on the cone $C(X)$.

\subsection{Supersymmetry and Holonomy}

In this subsection we wish to describe the amount of supersymmetry
preserved by the solutions $\ads_{p+2} \times X_d$ and $M_{p+1}\times
C(X)$, and the brane solution \eqref{eq:interpolatingbrane} that
interpolates between them.  In each case, the number of
supersymmetries is the number of linearly independent spinors
$\epsilon$ such that the supersymmetry variation of the gravitini
vanish in this background: $\delta_\epsilon \Psi =0$.  On $\ads_{p+2}
\times X_d$, such spinors are of the form $\chi \otimes \psi$ where
$\chi $ is a Killing spinor on $\ads_{p+2}$ and $\psi$ is a Killing
spinor on $X$, satisfying
\begin{equation}\label{eq:killingspinor}
\nabla^{(X)}_M \psi = \epsilon  \half \Gamma_M \psi
\end{equation}
where $\epsilon$ is either $+1$ or $-1$, depending on the field $F$.
For simplicity, and also for ease of comparison with the mathematical
literature on Killing spinors, we have rescaled the metric on the
Einstein manifold in such a way that the coefficient on the right-hand
side of equation \eqref{eq:killingspinor} is $\epsilon\half$.  The
number of supersymmetries of the $\ads_{p+2} \times X_d$ solution is
given by $\eN\,n_A\,n_X^\pm$ where $n_X^\pm$ is the number of Killing
spinors on $X$ with $\epsilon = \pm 1$, $n_A = 2^{\lfloor
\frac{p+2}{2} \rfloor}$ is the number of Killing spinors on
$\ads_{p+2}$ and $\eN=1$ for $\M$ theory and $\eN=2$ for Type II
strings.

If the dimension of $X$ is even, then there are as many solutions with
$\epsilon=1$ as with $\epsilon=-1$, whereas if $\dim X$ is odd, then
changing the sign of $\epsilon$ corresponds to reversing the
orientation of $X$.  Thus for odd-dimensional $X$, the number of
solutions depends on the orientation.  For example, for $d=7$, if
there are $n_X>0$ Killing spinors with one choice of orientation of
$X$, there will be no Killing spinors with the opposite choice of
orientation, unless $X$ is the round $7$-sphere, in which case $n_X=8$
with either choice of orientation \cite{DNP}.  We will see below that
this is very easy to understand in terms of the holonomy of the cone
$C(X)$.

On $M_{p+1}\times C(X)$, the supersymmetries are of the form $\chi
\otimes\psi$ where $\chi$ and $\psi$ are covariantly constant spinors
on Minkowski space $M_{p+1}$ and the cone $C(X)$, respectively. In
particular, $\psi$ satisfies
\begin{equation}\label{eq:parallelspinor}
\nabla^{(C)} \psi = 0~.
\end{equation}

The number of supersymmetries of the $M_{p+1}\times C(X)$ solution is
then given by $\eN\,n_M\,n_C$ where $n_C$ is the number of covariantly
constant spinors on the cone $C(X)$ and
$n_M=2^{\lfloor\frac{p+1}{2}\rfloor}$ is the number of parallel
spinors on $M_{p+1}$.

As shown in \cite{Baer}, there is a one to one correspondence between
Killing spinors on $X$ satisfying \eqref{eq:killingspinor} and
covariantly constant spinors on $C(X)$ satisfying
\eqref{eq:parallelspinor}.  Each covariantly constant spinor on $C(X)$
restricts to a Killing spinor on $X$ satisfying
\eqref{eq:killingspinor} for a particular orientation $\epsilon$ of
$X$.  If $\dim X$ is odd, so that $\dim C(X)$ is even, then the sign
of $\epsilon$ is also correlated to the chirality of the parallel
spinor on $C$.  If $d\equiv \dim X$ is even, so that $\dim C(X)$ is
odd, then the sign of $\epsilon$ depends on which one of the two
irreducible representations of the Clifford algebra $\Cl_{d+1}$ we use
to embed the unique spinor representation of $\Spin_{d+1}$.

In the round $7$-sphere compactification of $\M$ theory, there are $8$
solutions of \eqref{eq:killingspinor} with $\epsilon=1$ and $8$
solutions with $\epsilon=-1$, of which only $8$ have the right sign to
be Killing spinors, while $C(S^7)=\EE^8$ has $16$ covariantly constant
spinors. In this case, $n_X^\pm=8$, $n_C=16$, $n_A=4$ and $n_M=2$, so
there are $32$ supersymmetries on both $M\times C = \EE^{10,1}$ and
$\ads_4\times S^7$.  For other $X_7$, there are only solutions of
\eqref{eq:killingspinor} with {\em one\/} particular orientation
$\epsilon$, and parallel spinors on $C(X)$ restrict to spinors
satisfying \eqref{eq:killingspinor} with that choice of orientation.
This is because the spinors left invariant by any of the possible
holonomy groups which act irreducibly in eight dimensions all have the
same chirality.  On the other hand, as will be seen shortly, when $X$
is an Einstein manifold admitting Killing spinors and has dimension
$4k+1$, both orientations give the same numbers of supersymmetries.
For Type IIB compactifications on $5$-manifolds, this conflicts with a
claim made in \cite{DuffUntwisted}.  At any rate, we are interested in 
the supersymmetric cases, so we choose $\epsilon = +1$.

For $\M2$-, $\D3$- and $\M5$-branes, $n_A$ = $4,4,8$ respectively,
whereas $n_M$ = $2,4,8$. Using our result for the numbers of
asymptotic and near-horizon supersymmetries we find that for $\M2$-
and $\D3$-branes the number of supersymmetries of the near-horizon
geometry $\ads_{p+2} \times X_d$ is {\em twice\/} the number of
supersymmetries for the asymptotic conical geometry $M_{p+1}\times
C(X)$, unless $X$ is a round sphere, in which case the number of
supersymmetries is the same in both cases.  Applying this to the
$\M5$-brane gives the same number of supersymmetries asymptotically
and near the horizon, but in this case, as we shall see, the only
smooth spaces $X$ that admit Killing spinors are $S^4$ and $\RR P^4$.
In each of these cases, the number of supersymmetries can be further
reduced by orbifolding.  In particular, for the $\M5$-brane with
non-spherical near-horizon geometry, the asymptotic space is actually
an orbifold $M_6 \times \RR^5/\Gamma$ and in this case the
near-horizon limit has twice the number of supersymmetries as the
asymptotic region.

As an example, consider the squashed $7$-sphere of \cite{DLPS}, which
has $n_X^+=1$, and $n_X^-=0$, so that $\nu = \frac18$ or $\nu=0$ for
the squashed $7$-sphere compactification, depending on the  
orientation.  The cone has one parallel
spinor, so the near-horizon geometry has $\nu = \frac18$ or $\nu=0$,
while the asymptotic conical geometry has $\nu=\frac1{16}$.

In summary, the amount of supersymmetry on the brane and in the
near-horizon geometry is determined by the number of parallel spinors
on the cone $C(X)$, and this can now be analysed group-theoretically.

Assume that the base of the cone, $X$, is simply-connected.  Its cone
$C$ will be simply-connected also, since $X$ and $C$ are homotopy
equivalent.  Simply-connected manifolds admitting parallel spinors are
classified by their holonomy group \cite{Wang}.  Because $C$ is
Ricci-flat, we know that it cannot be a locally symmetric space.
Moreover, by a theorem of Gallot \cite{Gallot}, the holonomy group
acts on $C$ irreducibly unless $C$ is flat, in which case $X$ is the
round sphere. Therefore, we need only consider irreducible holonomy
groups of manifolds which are not locally symmetric.  In other words,
those in Berger's table (see, e.g., \cite{Besse,Salamon}).

Of those, the ones which admit parallel spinors are given in the
following table, which also lists the number $n$ (or $(n_L,n_R)$ in
even dimension) of linearly independent parallel spinors.

\begin{table}[h!]
\centering
\setlength{\extrarowheight}{3pt}
\begin{tabular}{|>{$}c<{$}|>{$}c<{$}|c|>{$}c<{$}|}\hline
\dim & \text{Holonomy} & Geometry& n\\[3pt]
\hline
\hline
4k+2 & \SU_{2k+1} & Calabi--Yau & (1,1)\\
4k & \SU_{2k} & Calabi--Yau & (2,0)\\
4k & \Sp_k & hyperk\"ahler & (k+1,0)\\
7 & G_2 & exceptional & 1\\
8 & \Spin_7  & exceptional & (1,0)\\[3pt]
\hline
\end{tabular}
\vspace{8pt}
\caption{Manifolds admitting parallel spinors}
\label{tab:berger}
\end{table}

If $X$, and hence $C$, were not simply-connected, then we could still
use part of the above analysis.  Gallot's theorem still applies and
the existence of parallel spinors constrains the {\em restricted\/}
holonomy group of the manifold to be contained in the above Table.
However a spinor which is invariant under the restricted holonomy
group need not be parallel, because it may still transform
nontrivially under parallel transport along noncontractible loops.
Therefore the number of parallel spinors in $C$ will be at most the
number shown in Table \ref{tab:berger}.

\section{Near-horizon Geometries of Cone Branes}

In this section we will discuss the relation between the geometry of
$X$ and the holonomy of its cone $C$.  We will do this in some
generality, and give the relation between the geometry of $X$ and the
number of supersymmetries preserved by a $\ads \times X$ solution,
when one exists, and this characterises the near-horizon geometries of
the supersymmetric cone branes discussed in the previous section.

On every cone there is a privileged vector field $\xi = r\d_r$, called
the {\em Euler vector\/}, which generates the rescaling
diffeomorphisms.  Moreover on any manifold of reduced holonomy, the
holonomy principle guarantees the existence of certain parallel
tensors, corresponding to the singlets under the holonomy group in the
tensor products of the holonomy representation.  Therefore on a cone
of reduced holonomy we will be able to build interesting geometric
structures out of these parallel tensors and the Euler vector.  These
structures will in turn induce interesting geometric structures on
$X$, which we identify with $\{1\} \times X \subset C$.  These
geometric structures are summarised in the following table, which also
lists the numbers $(n_+,n_-)$ of Killing spinors; that is, solutions
of equation \eqref{eq:killingspinor}, with $n_\pm$ the number of
solutions with $\epsilon =\pm 1$.

\begin{table}[h!]
\centering
\setlength{\extrarowheight}{3pt}
\begin{tabular}{|>{$}c<{$}|>{$}c<{$}|c|>{$}c<{$}|}\hline
\dim X & \text{Holonomy of $C$} & Geometry of $X$ & (n_+, n_-)\\
\hline
\hline
d & \{1\} & round sphere & (2^{\lfloor d/2\rfloor},2^{\lfloor
d/2\rfloor})\\
4k-1 &\Sp_k& $3$-Sasaki & (k+1,0)\\
4k-1 &\SU_{2k} & Sasaki--Einstein & (2,0)\\
4k+1 &\SU_{2k+1}& Sasaki--Einstein & (1,1)\\
6 & G_2 & nearly K\"ahler & (1,1)\\
7 & \Spin_7 & weak $G_2$ holonomy & (1,0)\\[3pt]
\hline
\end{tabular}
\vspace{8pt}
\caption{Manifolds admitting real Killing spinors}
\label{tab:geometries}
\end{table}

We now proceed to describe these geometries in detail.

\subsection{Cones with $\Spin_7$ Holonomy}

Any eight-dimensional manifold with $\Spin_7$ holonomy possesses a
parallel self-dual $4$-form $\Omega$, known as the Cayley form.
Contracting the Euler vector $\xi$ into the Cayley form gives a
$3$-form on $C$ which restricts to a $3$-form $\phi$ on $X$:
\begin{equation*}
\phi(u,v,w) \equiv \Omega(\xi,u,v,w)~,
\end{equation*}
for any vectors $u,v,w$ tangent to $X$.  In fact, one can write the
Cayley form restricted to $X\subset C$ as
\begin{equation*}
\Omega = dr \wedge \phi + \star \phi~,
\end{equation*}
with $\star$ the Hodge operator on $X$.  Notice that $dr$ is a
$1$-form on $C$, and by its restriction to $X$ we simply mean
evaluating it at $X\subset C$ on vector fields tangent to $C$.  In
other words, $dr$ acts both on vectors tangent and normal to $X$. From
the fact that $\Omega$ is parallel in $C$, it follows that, in $X$,
$\phi$ obeys
\begin{equation*}
\nabla \phi = \star\phi~.
\end{equation*}
This condition says that $X$ has {\em weak $G_2$ holonomy\/}
\cite{Gray-WH}, and we say that $\phi$ defines a {\em nearly parallel
$G_2$ structure\/}.  In fact, as proven in \cite{Baer} a manifold has
weak $G_2$ holonomy if and only if its metric cone has holonomy
contained in $\Spin_7$.  In the early Kaluza--Klein literature it
would have been said that $X$ has $G_2$ Weyl holonomy, but we will not
follow this nomenclature.

\subsection{Cones with $G_2$ holonomy}

On any seven-dimensional manifold with $G_2$ holonomy there is a
parallel $3$-form $\Phi$, known as the associative form.  Contracting
the Euler vector into the associative form yields a $2$-form on $C$
which restricts to a $2$-form $\omega$ on $X$:
\begin{equation*}
\omega(u,v) \equiv \Phi(\xi,u,v)~,
\end{equation*}
for any vectors $u,v$ tangent to $X$.  We can define a $(1,1)$-tensor
$J$ on $X$ as follows:
\begin{equation*}
\langle J(u),v \rangle \equiv \omega(u,v)~.
\end{equation*}
It is possible to show \cite{Baer} that $J$ is an orthogonal almost
complex structure:
\begin{equation*}
J^2 = -\1\qquad\text{and}\qquad \langle J(u),J(v)\rangle = \langle
u,v\rangle~,
\end{equation*}
whence $X$ is an almost hermitian manifold.  From the fact that $\Phi$
is parallel on $C$, it follows that on $X$,
\begin{equation*}
\nabla_v J(v) = 0\qquad\text{for any $v$;}
\end{equation*}
although $\nabla J \neq 0$.  Moreover $J$ is not integrable.  This
means that $X$ is a non-K\"ahler {\em nearly K\"ahler\/} manifold
\cite{GrayNK}.

In fact, the converse is also true and a six-dimensional manifold is
non-K\"ahler nearly K\"ahler if and only if its cone has holonomy
contained in $G_2$ \cite{Baer}.  A different proof that a
six-dimensional manifold admits Killing spinors if and only if it is
non-K\"ahler nearly K\"ahler appeared earlier in \cite{Grunewald}.

Nearly K\"ahler $6$-manifolds share many properties with Calabi--Yau
$3$-folds.  For example, there is a natural $3$-form defined by
contracting the Euler vector $\xi$ with the coassociative $4$-form
$\star\Phi$ on the cone.  In particular, nearly K\"ahler $6$-manifolds
have vanishing first Chern class \cite{GrayNK2}.  For example, $S^6
\cong G_2/\SU_3$ is nearly K\"ahler, but cannot be K\"ahler because
$H^2(S^6)$ is trivial.

\subsubsection{Scholium on Almost Hermitian Manifolds}

The notion of nearly K\"ahler manifolds should not be confused with
the notion of an {\em almost K\"ahler\/} manifold, which simply means
an almost hermitian manifold whose associated $2$-form $\omega$ is
closed; or in other words, a symplectic manifold with a compatible
almost complex structure.  Unfortunately, given the many different
generalisations of the notion of K\"ahler manifolds, there is a large
possibility for confusion, so we digress momentarily to settle the
notation once and for all.  As shown in \cite{AlmostHermitian}, there
are 16 classes of almost hermitian manifolds, the class of K\"ahler
manifolds being but one of them.  The classes are defined in the
following way.

One can define almost hermitian geometry in terms of $G$-structures.
Just like a Riemannian metric $g$ on an $2n$-dimensional manifold $X$
allows us to reduce the structure group of the tangent bundle from
$\GL_{2n}\RR$ to $\O_{2n}$, an almost hermitian structure $(g,J)$
allows us to reduce the group further to $\U_n$.  This means that
tensor bundles can be consistently decomposed into $\U_n$-irreducible
sub-bundles.  Let $\omega$ be the associated $2$-form.  Its covariant
derivative $\nabla\omega$ is a section through a sub-bundle $W \subset
T^*\otimes \bigwedge^2T^*$, with $T^*$ the cotangent bundle of $X$.
$W$ is not irreducible under $\U_n$, but rather decomposes into four
irreducible components $W = \bigoplus_{i=1}^4 W_i$.  There are
therefore sixteen $\U_n$-invariant sub-bundles (not necessarily
irreducible) in $W$: $\{0\}$, $W_i$, $W_i\oplus W_j$ ($i\neq
j$),\ldots, $W$.  The sixteen classes of hermitian manifolds
correspond to these sixteen sub-bundles: a manifold $X$ belonging to
the class corresponding to the sub-bundle of $W$ to which
$\nabla\omega$ belongs.

Clearly $\nabla\omega=0$ corresponds to the class of {\em K\"ahler\/}
manifolds; but there are other classes of manifolds which are close to
K\"ahler in some sense.  The class of non-K\"ahler {\em nearly
K\"ahler\/} manifolds can be shown to be the one for which $\nabla
\omega = \tfrac13 d\omega \neq 0$.  The class of {\em almost
K\"ahler\/} manifolds consists of those for which $d\omega = 0$.  To
make matters even more confusing there exist also {\em
semi-K\"ahler\/} manifolds for which $d\star\omega=0$ and {\em
quasi-K\"ahler\/} manifolds which have a more complicated definition
which we will not need.  For details the reader is referred to
\cite{AlmostHermitian}.

\subsection{Calabi--Yau Cones}

The class of Calabi--Yau $n$-folds is the class of compact Ricci-flat
K\"ahler $n$-folds.  Let us concentrate first on the K\"ahler
condition.  Any K\"ahler manifold has a parallel $2$-form, the
K\"ahler form, and a corresponding orthogonal complex structure $I$
which is also parallel.  Together with the Euler vector, we can
therefore build two objects: a $1$-form obtained by contracting the
Euler vector into the K\"ahler form, and a vector obtained by acting
with the complex structure on the Euler vector.  The vector is clearly
tangent to $X$ and restricts there to a vector $\chi$, whereas the
$1$-form restricts to a $1$-form $\theta$ on $X$ which is naturally
dual to $\chi$:
\begin{equation*}
\chi = I(\xi) \qquad\text{and}\qquad \theta = \langle \chi,-\rangle~.
\end{equation*}
It follows from the definition that $\chi$ has unit norm and is a
Killing vector.  Furthermore, the covariant derivative of $\chi$
defines a $(1,1)$-tensor $T \equiv \nabla\chi$ in $X$,
\begin{equation}\label{eq:SasakiTdef}
T(v) = \nabla_v\chi~,
\end{equation}
which, because $I$ is parallel in $C$, satisfies the following
identity:
\begin{equation}\label{eq:SasakiTid}
\nabla_u T(v) = \theta(v)\,u - \langle u,v\rangle\,\chi~.
\end{equation}
The triple $(\chi,\theta,T)$ defines a Sasaki structure on $X$.  More
precisely, a triple $(\chi,\theta,T)$ on an odd-dimensional Riemannian
manifold $X$, where $\chi$ is a unit norm Killing vector, $\theta$ its
dual $1$-form, and $T = \nabla\chi$ obeys equation
\eqref{eq:SasakiTid}, is called a {\em Sasaki\/} structure on $X$
\cite{Sasaki}, and $X$ is called {\em Sasakian\/}. Equivalently
\cite{Baer}, a manifold $X$ is Sasakian if and only if its metric cone
$C(X)$ is K\"ahler.

If in addition $C(X)$ is Ricci-flat (hence Calabi--Yau), then $X$ is
Einstein.  We say that $X$ is {\em Sasaki--Einstein\/}.  Therefore a
manifold is Sasaki--Einstein if and only if its metric cone is
Calabi--Yau \cite{FK5,FK7-2,Baer}.

This means, in particular, that a Sasaki--Einstein manifold $X$ of
dimension $2n+1$ also possesses two distinguished $n$-forms obtained
by contracting the Euler vector $\xi$ into the real and imaginary
parts of the complex volume ($n{+}1$)-form on $C(X)$.

The Killing vector $\chi$ in a $(2n+1)$-dimensional Sasakian manifold
$X$ defines a foliation whose leaves are the integral curves of
$\chi$.  The manifold is called {\em regular\/} if these curves are
closed and have the same length.  If this is the case, $\chi$ defines
a $\U_1$ action on $X$ and $X$ can be understood as a circle bundle
over the orbit space, a $2n$-dimensional manifold $M$, which can be
shown to be K\"ahler.  Moreover if $X$ is Sasaki--Einstein then $M$ is
K\"ahler--Einstein.  Regularity is a very stringent condition which is
not satisfied by most Sasaki(--Einstein) manifolds.  When the integral
curves of $\chi$ are closed but of different lengths, the orbit space
is an orbifold which is everywhere smooth except at a finite number of
points \cite{FK5,BoGa}.

\subsection{Hyperk\"ahler Cones}

In a hyperk\"ahler manifold we have a parallel quaternionic structure
consisting of three orthogonal complex structures $I,J,K$ satisfying
the quaternion algebra, as well as their corresponding K\"ahler forms:
$\omega_I, \omega_J, \omega_K$.  From the discussion above, on a cone
$C(X)$ each complex structure gives rise to one Sasaki structure on
$X$.  Moreover the fact that the three complex structures on $C(X)$
satisfy the quaternion algebra means that the Killing vectors in the
three Sasaki structures are orthonormal and obey an $\su_2$ Lie
algebra.  Three Sasaki structures satisfying these conditions define a
{\em $3$-Sasaki\/} structure on $X$ (see \cite{BoGa} for the latest
word on $3$-Sasakian manifolds).  Equivalently, one can prove the
converse and show that a manifold is $3$-Sasakian if and only if its
cone is hyperk\"ahler \cite{FK7-2,Baer,BGM}.

The three Killing vectors define a foliation of the $3$-Sasakian
manifold $X$.  We say that $X$ is {\em regular\/} if the foliation
fibres.  This means that $X$ is an $\SU_2$ or $\SO_3$ bundle over a
quaternionic K\"ahler space $Q$.  Equivalently, $X$ is a circle bundle
over the twistor space $Z$ of $Q$.  If $X$ is not regular, but the
Killing vectors are complete, then the orbit space defines a
quaternionic K\"ahler orbifold \cite{BoGa}.

\section{Examples}

In this section we list the known near-horizon geometries for
different types of cone branes.  The homogeneous examples are all
known from the early days of Kaluza--Klein supergravity, but we do
list some non-homogeneous examples as well.  For each type of cone
brane we discuss the possible smooth near-horizon geometries and where
applicable we discuss orbifolds.

\subsection{$\M2$ Cone Branes}

The transverse space to an $\M2$ cone brane is the metric cone $C_8$
over a $7$-dimensional manifold $X_7$.  From Table
\ref{tab:geometries} one can read off the following possibilities for
simply-connected $X_7$, which are listed in Table \ref{tab:se7en}.  In
the non-simply-connected case the number of Killing spinors is at most
the one shown in the table.

\begin{table}[h!]
\centering
\setlength{\extrarowheight}{3pt}
\begin{tabular}{|c|c|>{$}c<{$}|}\hline
\text{Geometry of $C_8$} & Geometry of $X_7$ & (n_+, n_-)\\
\hline
\hline
flat & round $S^7$ & (8,8)\\
hyperk\"ahler & $3$-Sasaki & (3,0)\\
Calabi--Yau & Sasaki--Einstein & (2,0)\\
$\Spin_7$ holonomy & weak $G_2$ holonomy & (1,0)\\[3pt]
\hline
\end{tabular}
\vspace{8pt}
\caption{Seven-dimensional near-horizon geometries}
\label{tab:se7en}
\end{table}

Besides the sphere and its quotients, we have the following types of
near-horizon geometries.

\subsubsection{$X_7$ is $3$-Sasakian}

We must distinguish between regular and non-regular $3$-Sasakian
$7$-manifolds.  The regular manifolds were classified in
\cite{FK7-2,BGM} and they all happen to be homogeneous spaces.
Therefore they have been known since the early days of Kaluza--Klein
supergravity \cite{CastellaniRomansWarner}.  In the present context
they have been discussed in \cite{CCDAFFT}.  Apart from $\Sp_2/\Sp_1
\cong S^7$ and $\Sp_2/\left(\Sp_1\times\ZZ_2\right)\cong \RP^7$, the
only other homogeneous regular example is
$\left(\SU_3\times\U_1\right)/\left(\U_1 \times \U_1\right)$, which is
called $N^{010}$ in \cite{CCDAFFT} and $N(1,1)$ in \cite{DNP}.

On the other hand there are an infinite number of different
non-regular $3$-Sasakian $7$-manifolds.  The topology of a
seven-dimensional $3$-Sasakian manifold $X_7$ is highly constrained.
First of all, Bochner's theorem implies that the first Betti number of
a $3$-Sasakian manifold vanishes: $b_1=0$.  In seven dimensions,
$b_6=0$ by Poincar\'e duality.  Furthermore, as shown in
\cite{GalickiSalamon,FKMS}, $b_3=b_4=0$, so that the only nonzero
Betti numbers are $b_2 = b_5$.  Recently an infinite family with
arbitrary $b_2$ has been constructed in \cite{BGMR,Bielawski}.  This
implies that there exist $3$-Sasakian $7$-manifolds of every possible
rational homotopy type allowed.  These examples are toric; that is,
they admit an action of a torus $\TT^2$ preserving the $3$-Sasakian
structure.  They are constructed via a $3$-Sasakian quotient
\cite{BGM3SQ} akin to the hyperk\"ahler quotient \cite{HKLR}.  Indeed,
the two quotients are related via the cone construction.  In other
words, suppose $X_7$ is a $3$-Sasakian manifold and $C_8$ is its
hyperk\"ahler cone.  Then if $C_8$ admits a triholomorphic action
{\em commuting with the Euler vector\/}, then the hyperk\"ahler
quotient of $C_8$ is a cone over the $3$-Sasaki quotient of $X_7$.
The cones over the toric $3$-Sasakian manifolds in
\cite{BGMR,Bielawski} form a subclass of the toric hyperk\"ahler
manifolds studied in \cite{BD} and which have been considered in the
context of intersecting branes in \cite{GGPT}.

All $3$-Sasakian manifolds have an infinitesimal $\su_2\cong\so_3$
isometry.  If the Killing vectors are complete, they integrate to an
action of $\SU_2$ or $\SO_3$.  In the regular case, the orbit space is
smooth; otherwise not.  In any case, the infinitesimal isometry is a
generic feature of these manifolds and one which will play a role in
Section \ref{sec:scft}, when we discuss the dual field theories.

\subsubsection{$X_7$ is Sasaki--Einstein}

There are many known Sasaki--Einstein $7$-manifolds.  The regular
examples can be understood as $\U_1$ bundles over K\"ahler--Einstein
$3$-folds \cite{PopeWarner}.  They have not been classified.  The
known results are summarised in the following table \cite{FKMS}.

\begin{table}[h!]
\centering
\setlength{\extrarowheight}{3pt}
\begin{tabular}{|>{$}c<{$}|>{$}c<{$}|}\hline
K_6 & X_7 \\
\hline
\hline
F(1,2|3) & N^{010} \cong
\left(\SU_3\times\SU_2\right)/\left(\SU_2\times\U_1\right)\\
\CP^1 \times \CP^1 \times \CP^1 & Q^{111} \cong
\left(\left(\SU_2\right)^3 \times \U_1\right)/
\left(\left(\U_1\right)^3 \times\U_1\right)\\
\CP^2 \times \CP^1 & M^{110}\cong
\left(\SU_3\times\SU_2\times\U_1\right)/
\left(\SU_2\times\U_1\times\U_1\right) \\
G(2|5) & V(2|5)\cong \left(\SO_5\times\U_1\right) /
\left(\SO_3\times\U_1\right)\\
P_k \times \CP^1 & M_k^7~(3\leq k \leq 8)\\[3pt]
\hline
\end{tabular}
\vspace{8pt}
\caption{Known regular Sasaki--Einstein $7$-manifolds $X_7$, as $\U_1$
bundles $X_7\to K_6$ over K\"ahler--Einstein $3$-folds}
\label{tab:regularSE7}
\end{table}

The first four examples are homogeneous and have therefore appeared in
the early Kaluza--Klein literature.  The first three are listed using
the nomenclature of \cite{CastellaniRomansWarner}.  In \cite{DNP,FKMS}
these spaces are called $N(1,1)$, $Q(1,1,1)$ and $M(3,2)$,
respectively.  Here $V(2|5)$ is the Stiefel manifold of orthonormal
oriented $2$-frames in $\RR^5$ and $G(2|5)$ is the Grassmannian of
oriented $2$-planes in $\RR^5$.  $F(1,2|3)$ is the complex flag
manifold in $\CC^3$ consisting of pairs $(\ell,\pi)$ where $\pi$ is a
complex plane in $\CC^3$ and $\ell \subset \pi$ is a complex line.
The last class of examples does not consist of homogeneous manifolds:
$P_k$ are the del Pezzo surfaces consisting of blowing up $k$ points
in general position on $\CP^2$.

In addition, there are two infinite families of homogeneous
non-regular Sasaki--Einstein $7$-manifolds: the $M^{ppr}$ of
\cite{WittenMpqr,CDAF} and the $Q^{ppp}$ of \cite{DAFvN}.  Of course
they are mentioned in \cite{CastellaniRomansWarner}.

\subsubsection{$X_7$ has Weak $G_2$ Holonomy}

The canonical example of a weak $G_2$ holonomy manifold which is not
Sasaki--Einstein is the squashed $7$-sphere, which is a homogenous
space $\left(\Sp_2\times\Sp_1\right)$\newline $/\left(\Sp_1\times\Sp_1\right)$.
It turns out that this generalises, and every $3$-Sasakian
$7$-manifold can be squashed to a manifold with weak $G_2$ holonomy
\cite{GalickiSalamon,FKMS}.  The squashing is done as follows.  A
$3$-Sasakian manifold $X_7$ is foliated by the action of the Sasakian
Killing vectors.  At any point $p\in X_7$, the Killing vectors are
tangent to the unique leaf $\eF_p$ of the foliation passing through
$p$.  The tangent space to $X$ at $p$ has an orthogonal decomposition
$T_pX = \eF_p \oplus \left(\eF_p\right)^\perp$.  The squashing of the
metric is done by introducing a parameter $t$ and rescaling the metric
on the leaves of the foliation by $t$.  We can then compute the Ricci
tensor as a function of $t$ and notice that there are two values for
which it is Einstein: one is the original $3$-Sasakian metric, and the
other gives rise to a weak $G_2$ holonomy structure.

By squashing the infinite toric family of \cite{BGMR,Bielawski} in
this way, one obtains a huge number of weak $G_2$ holonomy manifolds
which are not Sasaki--Einstein.  These spaces are not homogeneous
and hence have not been considered before in the Kaluza--Klein
supergravity context.  There are examples with arbitrary $b_2=b_5$ and
all other Betti numbers vanishing, since squashing does not change the
topology.

Another infinite dimensional family of weak $G_2$ holonomy manifolds
consists of the Aloff--Wallach spaces \cite{AloffWallach} $N(k,l)
\cong \SU_3/\U_1$, with $\U_1 \subset \SU_3$ defined by
\begin{equation*}
e^{i\theta} \mapsto
\begin{pmatrix}
e^{ik\theta} &  & \\
 & e^{i\ell\theta} & \\
 &  & e^{-i(k+\ell)\theta}
\end{pmatrix}~.
\end{equation*}
This family is remarkable because it contains an infinite number of
distinct homotopy types and even exotic pairs, i.e., homeomorphic
non-diffeomorphic pairs \cite{KreckStolz}.  For $k\neq 1$ or
$\ell\neq 1$ these spaces admit metrics with weak $G_2$ holonomy.
The Aloff--Wallach spaces $N(k,\ell)$ agree, allowing for some
redundancy in the notation, with the $N^{pqr}$ spaces of
\cite{CastellaniRomans}.  Finally, the other weak $G_2$ holonomy
manifold from the early Kaluza--Klein literature is the homogeneous
space $\SO_5/\SO_3^{\mathrm{max}}$ of \cite{CastellaniRomansWarner}.

In summary, all known examples are homogeneous, hence previously known
in the Kaluza--Klein context, except for the squashed toric
$3$-Sasakian manifolds.

\subsection{$\M5$ Cone Branes}

The transverse space to an $\M5$ cone brane is a five-dimensional cone
$C_5$ over a four-dimensional manifold $X_4$.  As proven in \cite{F4},
any complete connected $n$-dimensional spin manifold, with $n\leq4$,
admitting real Killing spinors is locally isometric to the round
$n$-sphere.  Therefore the only possible {\em smooth\/} near-horizon
geometry for an $\M5$ cone brane is either spherical $S^4$ or
elliptical $\RP^4$; these describe the near-horizon geometry of $\M5$
branes at a point in $\RR^5$ or on the orientifold $\RR^5/\ZZ_2$.
More generally we can consider orbifolds of $\RR^5\cong \RR \times
\CC^2$ which are of the form $\RR\times\CC^2/\Gamma$. Such an orbifold 
will preserve supersymmetry if $\Gamma$ is an ADE subgroup of the
hyperk\"ahler $\SU_2$ which acts on $\CC^2$.  This clearly induces a
near-horizon geometry of the form $S^4/\Gamma$ which has two singular
points induced from the orbifold fixed points of $\Gamma$ acting on
$\RR \times \CC^2$.  These orbifolds preserve half of the
supersymmetries obtained in the maximally supersymmetric case.

\subsection{$\D3$ Cone Branes}

The transverse space to a $\D3$ cone brane is a six-dimensional cone
$C_6$ over a five-dimensional manifold $X_5$.  From Table
\ref{tab:geometries} we see that there are two simply-connected
possibilities.  This means $X_5$ is a sphere or a Sasaki--Einstein
manifold.  Non-simply connected examples, e.g., $\RP^5$ or more
generally $S^5/\Gamma$, can be obtained by taking quotients.
There are an infinite number of smooth quotients $S^5/\Gamma$ of the
sphere which possess Killing spinors.  These can be determined as
follows \cite{S5/G}.

In his solution of the spherical space problem, Wolf classified all
the discrete subgroups $\Gamma\subset\SO_6$, for which $S^5/\Gamma$ is
smooth \cite{Wolf}.  Given one such subgroup, the spin structures on
$S^5/\Gamma$ are in one-to-one correspondence with the lifts of
$\Gamma$ to an isomorphic subgroup $\Tilde\Gamma \subset
\Spin_6\cong\SU_4$; that is, to a $\Tilde\Gamma \subset \Spin_6$ which
is mapped to $\Gamma$ isomorphically under the covering map $\Spin_6
\to \SO_6$.  Finally, for $S^5/\Gamma$ with spin structure given by
$\Tilde\Gamma$, the Killing spinors in $S^5/\Gamma$ are precisely the
$\Tilde\Gamma$-invariant spinors in $S^5$.  The results are as
follows.

There are two families of subgroups of $\SO_6$ for which $S^5/\Gamma$
is smooth: $\Gamma(n,a,b)$ and $\Gamma(m,r;n,s)$ described below.

Let $\Gamma(n,a,b)$, where $n,a,b\in\NN$ be the cyclic subgroup of
$\SO_6$ of order $n$ generated by the following element:
\begin{equation*}
\begin{pmatrix}
R\left(\tfrac{1}{n}\right) & & \\
& R\left(\tfrac{a}{n}\right) & \\
& & R\left(\tfrac{b}{n}\right)
\end{pmatrix}
\qquad\text{with}\qquad
\begin{matrix}
(a,n) = (b,n) = 1\\
1 \leq a,b < n
\end{matrix}~,
\end{equation*}
where $R(\theta)$ denotes the rotation matrix
\begin{equation*}
R(\theta) = \begin{pmatrix}
	    \phantom{-}\cos 2\pi\theta & \sin 2\pi\theta\\
            - \sin 2\pi\theta & \cos 2\pi\theta
	    \end{pmatrix}~,
\end{equation*}
and $(p,q)$ denotes the greatest common divisor, so that $(p,q)=1$
means that $p$ and $q$ are coprime.

Similarly let $\Gamma(m,r;n,s)$, with $m,n,r,s\in\NN$ and $n \equiv 0
\pmod{3}$, denote the subgroup of $\SO_6$ generated by elements $A$
and $B$ subject to the relations: $A^m = B^n = \1$ and $B\,A\,B^{-1} =
A^r$.  The generators can be written as
\begin{equation*}
A = \begin{pmatrix}
    R\left(\tfrac{1}{m}\right) & & \\
    & R\left(\tfrac{r}{m}\right) & \\
    & & R\left(\tfrac{r^2}{m}\right)
     \end{pmatrix}
\qquad\text{and}\qquad
B = \begin{pmatrix}
    0 & \1 & 0\\
    0 & 0 & \1\\
    R\left(\tfrac{3s}{n}\right) & 0 & 0
    \end{pmatrix}~,
\end{equation*}
where $\1$ is a $2\times 2$ identity matrix, and where $\left(
(r-1)n,m\right) = (s,\frac{n}{3}) = 1$ and $r\not\equiv r^3
\equiv 1\pmod{m}$ and $\1$ is the identity matrix of order 2.  This
subgroup has order $nm$.

It is not hard to show that $S^5/\Gamma(n,a,b)$ has precisely one
spin structure when $n$ is odd, and none for $n$ even.
Similarly $S^5/\Gamma(m,r;n,s)$ has precisely one spin structure when
both $n$ and $m$ are odd, and none otherwise.

Computing the $\Tilde\Gamma$-invariant spinors for each of the cases
above is an easy matter.  One sees that $S^5/\Gamma(n,a,b)$ is of type
$(n_+,n_-) = (1,1)$ whenever any one of the following equalities is
satisfied:
\begin{equation*}
a + b \pm 1 = n\qquad\text{or}\qquad a - b = \pm 1~,
\end{equation*}
and has no Killing spinors otherwise.  Similarly,
$S^5/\Gamma(m,r;n,s)$ has no Killing spinors for $n>3$; and
$S^5/\Gamma(m,r;3,s)$, $m$ odd, is of type $(n_+,n_-) = (1,1)$
whenever $r^2 + r +1 \equiv 0 \pmod{m}$ and has no Killing spinors
otherwise.

Notice that most of these spaces $S^5/\Gamma$ are not homogeneous.
It is likely that, as in the case of orbifolds \cite{KaSi-orbifolds},
the spectrum of the dual gauge theory associated with a near-horizon
geometry of the form $\ads_5\times S^5/\Gamma$ agrees with the
$\Tilde\Gamma$-invariant spectrum of $\eN{=}4$ supersymmetric
Yang--Mills.

\begin{table}[h!]
\centering
\setlength{\extrarowheight}{3pt}
\begin{tabular}{|c|c|>{$}c<{$}|}\hline
\text{Geometry of $C_6$} & Geometry of $X_5$ & (n_+, n_-)\\
\hline
\hline
flat & round $S^5$ & (4,4)\\
Calabi--Yau & Sasaki--Einstein & (1,1)\\[3pt]
\hline
\end{tabular}
\vspace{8pt}
\caption{Five-dimensional near-horizon geometries}
\label{tab:five}
\end{table}

\subsubsection{$X_5$ is Sasaki--Einstein}

Again we distinguish between regular and non-regular Sasaki--Einstein
manifolds.  The regular manifolds were classified in \cite{FK5}.
Regularity implies that $X_5$ is the total space of a $\U_1$ bundle
over a four-dimensional manifold $K_4$, which inherits a
K\"ahler--Einstein structure.  The results are summarised in Table
\ref{tab:regularSE5} below, where as in Table \ref{tab:regularSE7}
above, $P_k$ denotes the del Pezzo surfaces and $V(2|4)$ denotes the
Stiefel manifold of orthonormal oriented $2$-frames in $\RR^4$.  In
the table $\#$ denotes the operation of connected sum. For
completeness the non-simply-connected cases have also been
included. There are no known examples of Sasaki--Einstein
$5$-manifolds which are not regular.  (See Problem 8.1 in
\cite{BoGa}.)

\begin{table}[h!]
\centering
\setlength{\extrarowheight}{3pt}
\begin{tabular}{|>{$}c<{$}|>{$}c<{$}|}\hline
K_4 & X_5 \\
\hline
\hline
\CP^2 & S^5\\
\CP^2 & S^5/\ZZ_3\\
\CP^1 \times \CP^1 & V(2|4) \cong S^2 \times S^3\\
\CP^1 \times \CP^1 & V(2|4)/\ZZ_2\\
P_k~,~3\leq k \leq 8 & S_k \cong \#^k\left(S^2 \times
S^3\right)\\[3pt]
\hline
\end{tabular}
\vspace{8pt}
\caption{Regular Sasaki--Einstein $5$-manifolds $X_5$, as $\U_1$
bundles $X_5\to K_4$ over K\"ahler--Einstein surfaces}
\label{tab:regularSE5}
\end{table}

The homogeneous cases were known from the early Kaluza--Klein
literature.  In particular, $V(2|4) \cong
\left(\SU_2\times\SU_2\right)/\U_1$ is called $T^{1,1}$ in
\cite{Romans5}.  This solution $V(2|4)$ has also been discussed
recently in \cite{DuffUntwisted} where it was denoted $ Q(1,1)$ and
claimed as new.  These authors moreover claim that $Q(1,1)$ has
type $(n_+,n_-)=(2,0)$.  Since this space is simply-connected, its
cone is also simply connected. If $(n_+,n_-)=(2,0)$ the cone would
have to possess two covariantly constant spinors of one chirality and
none of the other.  Since the cone is six-dimensional the only
possible holonomy group is $\SU_3$ which has two parallel spinors, one
of each chirality, in contradiction with the claim in
\cite{DuffUntwisted}.  This claim seems to have propagated to
\cite{Kehagias}, where the dual SCFT is claimed to have $\eN{=}2$
supersymmetry.  However, as pointed out correctly in
\cite{KlebanovWitten}, the theory has only $\eN{=}1$ supersymmetry as
expected from our geometric considerations.

\subsection{Six Dimensions}

The remaining case to consider is that of a seven-dimensional cone
$C_7$ over a six-dimensional manifold $X_6$. In this case, there is no
known maximally supersymmetric compactification on $S^6$ to an anti-de
Sitter space, so the relevance of this case is unclear.  From Table
\ref{tab:geometries} we see that there are again two possibilities:
the sphere $S^6$ and the nearly K\"ahler non-K\"ahler manifolds.

\begin{table}[h!]
\centering
\setlength{\extrarowheight}{3pt}
\begin{tabular}{|c|c|>{$}c<{$}|}\hline
\text{Geometry of $C_7$} & Geometry of $X_6$ & (n_+, n_-)\\
\hline
\hline
flat & round $S^6$ & (8,8)\\
$G_2$ holonomy & nearly K\"ahler & (1,1)\\[3pt]
\hline
\end{tabular}
\vspace{8pt}
\caption{Six-dimensional near-horizon geometries}
\label{tab:six}
\end{table}

\subsubsection{$X$ Is Nearly K\"ahler}

As we mentioned above $S^6 \cong G_2/\SU_3$ is nearly K\"ahler
non-K\"ahler.  There are other examples.  First we have $\CP^3$.
Although $\CP^3$ possesses a K\"ahler metric, the $\SO_5$-invariant
one it inherits from the isomorphism $\CP^3 \cong \SO_5/\U_2$ is
non-K\"ahler nearly K\"ahler.  The same occurs for the complex flag
space $F(1,2|3)$ and the $\U_3$-invariant metric coming from $F(1,2|3)
\cong \U_3/ \left(\U_1 \times \U_1 \times \U_1\right)$
\cite{Gray-3S}.  The Lie group $\Spin_4 \cong S^3 \times S^3$ is
nearly K\"ahler with the metric inherited from a $3$-symmetric
structure \cite{Grunewald}.  It cannot be K\"ahler because
$H^2(S^3\times S^3) = 0$.  Similarly the homogeneous spaces
$\SO_5/\left(\U_1\times\SO_3\right)$, $\SO_6/\U_3$ and $\Sp_2/\U_2$
with their natural homogeneous metrics are nearly K\"ahler
non-K\"ahler \cite{Gray-KSAH}.

\section{Near-Horizon Geometry And Superconformal \\   Symmetry}
\label{sec:scft}

According to the conjecture in \cite{Malda} superstring or $\M$ theory
compactified on the near-horizon geometry $\ads_{p+2}\times X_d$ of (a
large number of) parallel $p$-branes, is dual to (the `t~Hooft limit
of) a superconformal field theory on the worldvolume of the brane,
which is the boundary of the anti-de~Sitter space.  The symmetry
algebras of the conformal field theory and the anti-de~Sitter string
theory must be the same superalgebra.  The anti-de~Sitter symmetry
algebra is the super-isometry algebra of the background; it has a
bosonic subalgebra which is the isometry algebra $\so_{p+1,2} \times
\fg$ of  $\ads_{p+2}\times X_d$, where $\fg$ is the isometry algebra
of $X_d$, and there is a fermionic generator for each Killing spinor,
generating the corresponding supersymmetry transformation.  The same
superalgebra acts as the superconformal algebra for the worldvolume
theory in $M_{p+1}$, where $\so_{p+1,2}$ is the conformal algebra in
$p+1$ dimensions, $\fg$ is a global symmetry which is the
$\R$-symmetry, or the product of the $\R$-symmetry with an algebra
that does not act on the fermionic generators, and the fermionic
generators consist of the usual supercharges in $p+1$ dimensions,
together with the special supersymmetry generators.

In the dimensions of interest here, these algebras were classified by
Nahm \cite{Nahm}, although the case $p=3$ had been considered
previously in \cite{HLS}.  Each such superalgebra is characterised by
its bosonic subalgebra together with the representation that the
supercharges are in.  This means that for the anti-de~Sitter vacua of
interest here, the bosonic subalgebra should contain $\so_{p+1,2}
\times \fg$, where $\fg$ is the isometry algebra of $X_d$.  This
information, together with Nahm's classification, is enough to
identify the superconformal algebra of the dual theory, provided that
the supercharges transform correctly.  In this section we will show
that the geometric structures we have listed previously are precisely
enough to identify the correct superconformal algebra in all cases.
We do this by identifying the generic isometries of the near-horizon
geometries and by investigating how the Killing spinors transform
under them.

\subsection{$d{=}3$ SCFTs Dual To $\M2$-Branes}

The superconformal algebra corresponding to an $\eN$-extended
three-dimen-sional superconformal theory is $\osp_{4|\eN}$ which has
bosonic subalgebra
\begin{equation*}
\so_{3,2} \times \so_\eN~,
\end{equation*}
with the supercharges transforming according to the
$(\repre{4},\repre{\eN})$ \cite{Nahm}.  Notice that for $\eN=8$, there
are three possible eight-dimensional representations: by $\repre{8}$
we mean one of the spinorial irreducible representations.

These theories are dual to $\M2$-branes at a conical singularity of a
Ricci-flat space with holonomy contained in $\Spin_7$.  We should
therefore identify the bosonic subalgebra as isometries of the
near-horizon geometry.  The $\so_{3,2}$ factor is the isometry of
$\ads_4$, while the $\so_\eN$ factor is a subalgebra of the isometry
algebra of $X_7$.  For a supersymmetric brane configuration, the cone
can have holonomy $\{1\}$, $\Sp_2$, $\SU_4$ and $\Spin_7$,
corresponding to near-horizon geometries preserving $32$, $12$, $8$
and $4$ supercharges respectively. In terms of three-dimensional
superconformal symmetries, each of which having four real components,
these cases thus correspond to $\eN{=}8,3,2,1$, respectively.  We now
treat each case separately.

\subsubsection{$\eN{=}8$}

This is the maximally supersymmetric case, in which the near-horizon
geometry is spherical.  The isometry algebra of the round $7$-sphere
is indeed $\so_8$.  It is clear that the Killing spinors are in the
$\repre{8}_s$ or $\repre{8}_c$ of $\so_8$.  Therefore one recovers, as
was done originally in \cite{Malda}, the correct superconformal
algebra.

\subsubsection{$\eN{=}3$}

The corresponding near-horizon geometry in this case has a cone base
which is a $3$-Sasakian $7$-manifold $X_7$.  As we saw above, such a
manifold possesses three orthonormal Killing vectors which generate an
action of $\su_2\cong\so_3$.  We need to check that the Killing
spinors transform as the $\repre{3}$ of this $\so_3$.  This is proven
in Theorem~3.1 in \cite{Moroianu}.  Alternatively, we can see this in
the hyperk\"ahler cone.  Each Killing spinor in the base corresponds
to a parallel spinor in the cone.  From the explicit construction of
the Killing vectors in the $3$-Sasaki structure at the base of the
cone, we see that their Lie algebra is isomorphic to the $\sp_1$
factor in the maximal subalgebra $\sp_1\times\sp_2 \subset\so_8$.
Under this subalgebra, the spinor representation $\repre{8}_s$
decomposes as
\begin{equation*}
\repre{8}_s \to (\repre{3},\repre{1}) \oplus (\repre{1},\repre{5})~.
\end{equation*}
Thus we see that the three parallel spinors transform as the
$\repre{3}$ of $\sp_1 \cong \so_3$.

\subsubsection{$\eN{=}2$}

In this case the base of the cone is a Sasaki--Einstein $7$-manifold
$X_7$, and as we saw above it has a unit norm Killing spinor $\chi$.
According to Theorem~2.2 in \cite{Moroianu}, the Killing spinors
transform according to the real two-dimensional irreducible
representation $\repre{2}$ of the $\so_2$ action of $\chi$.
Alternatively we can again work in the cone.  We notice that the
abelian algebra generated by $\chi$ corresponds to the $\u_1$ factor
of the maximal subalgebra $\u_1 \times \su_4 \cong \u_4 \subset
\so_8$.  Under $\so_8 \supset \u_1\times\su_4$, the spinor
representation $\repre{8}_v$ breaks up as
\begin{equation*}
\repre{8}_s \to (\repre{2},\repre{1}) \oplus (\repre{1},\repre{6})~,
\end{equation*}
so that the parallel spinors do indeed transform as the $\repre{2}$ of
$\u_1\cong\so_2$.

\subsubsection{$\eN{=}1$}

This case corresponds to the cones whose base is a $7$-manifold with
weak $G_2$ holonomy.  Generically the nearly parallel $G_2$ structure
admits no infinitesimal automorphisms, although of course we gave
plenty of examples above of such manifolds with a large isometry
group.  The Lie algebra of this group acts trivially on the Killing
spinor, however.  The automorphism groups of nearly parallel $G_2$
structures have been studied in \cite{FKMS}.  As an example, if $X_7$
is the squashed $7$-sphere with isometry $\sp_2\times \sp_1$, the
symmetry algebra is $\osp_{4|1}\times \sp_2\times \sp_1 $, with
trivial $\R$-symmetry and a symmetry $\sp_2\times\sp_1$ under which
the supercharges are singlets.

\subsection{$d{=}4$ SCFTs Dual To $\D3$-Branes}

From the existence of supersymmetric solutions of IIB theory
describing parallel $\D3$-branes sitting at conical singularities in
Ricci-flat six-dimensional cones with holonomy contained in $\SU_3$,
we expect a correspondence between $d{=}4$ SCFTs and IIB theory in
$\ads_5\times X_5$ with $X_5$ a Sasaki--Einstein $5$-manifold.  The
superalgebras governing the superconformal field theories
\cite{HLS,Nahm} have bosonic subalgebras
\begin{equation*}
\begin{cases}
\so_{4,2} \times \u_\eN & \text{for $\eN\neq 4$}\\
\so_{4,2} \times \su_4& \text{for $\eN{=}4$,}
\end{cases}
\end{equation*}
with the fermions in the representation
$\real{(\repre{4},\repre{\eN})}$, where we are using the notation
$\real{R}$ to mean the underlying real representation in $R \oplus
R^*$; in other words, $\real{R}\otimes_\RR \CC \cong R \oplus R^*$.
The $\so_{4,2}$ factor is the isometry algebra of the anti-de~Sitter
space, and the $\u_\eN$ (or $\su_4$) factor should coincide with the
isometries of $X_5$.  Moreover the isometries should act on the
Killing spinors according to the above representation.

\subsubsection{$\eN{=}4$}

This corresponds to the maximally supersymmetric case in four
dimensions, where $X_5$ is the round sphere.  The isometry group of the
sphere has Lie algebra $\so_6 \cong \su_4$, which was identified in
\cite{Malda} with the $\R$-symmetry group of the $\eN{=}4$ SCFT.

\subsubsection{$\eN{=}2$}

This case corresponds to $X_5\cong S^5/\Gamma$ where $\Gamma$ is any
of the ADE subgroups of the `hyperk\"ahler' $\Sp_1$ subgroup of
$\Sp_1\cdot\Sp_1 \cong \SO_4 \subset \SO_4 \times \SO_2 \subset
\SO_6$.  The centraliser of its Lie algebra $\sp_1$ in $\so_6$ is
therefore $\sp_1\times\so_2 \cong \su_2 \times\u_1 \cong \u_2$, as
expected.  We need to verify that the action on the Killing spinors is
the right one.  For the sphere, all spinors are Killing, so that they
transform under $\so_6 \cong \su_4$ as $\real{\repre{4}}$.  Under the
maximal $\su_2\times\su_2\times\so_2\subset\so_6$ subgroup, we have
the following branchings
\begin{equation*}
\repre{4} \to (\repre{1},\repre{2})_{+1} \oplus
(\repre{2},\repre{1})_{-1}
\qquad\text{and}\qquad
\repre{4}^* \to (\repre{1},\repre{2})_{-1} \oplus
(\repre{2},\repre{1})_{+1}~.
\end{equation*}
The $\Gamma$-invariant spinors then transform according to
$\real{\repre{2}_{+1}}$, which is precisely the $\real{\repre{2}}$ of
$\u_2$.

\subsubsection{$\eN{=}1$}

In this case the cone base $X_5$ is a Sasaki--Einstein $5$-manifold.
These manifolds possess a Killing spinor which generates a $\u_1$
subalgebra of the isometry algebra.  To see how the $\u_1$ acts on the
supersymmetries it is again convenient to work in the cone.  The
Sasakian Killing vector generates the $\u_1$ subalgebra of the maximal
subalgebra $\u_1\times\su_3\cong \u_3 \subset \so_6$.  Under this
subgroup the spinor representation $\real{\repre{4}}$ of
$\so_6\cong\su_4$ breaks up as
\begin{equation*}
\real{\repre{4}} \to \real{\repre{3}_{+1}} \oplus
\real{\repre{1}_{-3}}~.
\end{equation*}
Therefore the parallel spinors on the cone transform under $\u_1$ as
a real two-dimensional representation $\real{(+3) \oplus (-3)}$, in
agreement with the trasnformation properties of the supercharges.

\subsection{$d{=}6$ SCFTs Dual To $\M5$-Branes}

If we consider an $\M5$ cone brane, we might obtain $d{=}6$ chiral
SCFTs dual to $\M$ theory compactified on Einstein $4$-manifolds
admitting Killing spinors. There are only two cases: $\eN{=}1,2$,
although strictly speaking for $\eN{=}1$ we will have to consider
orbifolds.  The corresponding superconformal algebras \cite{Nahm} have
bosonic subalgebra
\begin{equation*}
\so_{6,2} \times \sp_\eN~,
\end{equation*}
with the fermions in the real representation
$[(\repre{8},\repre{2\eN})]$, where we have introduced the notation
$[R]$ to mean the underlying real representation of a complex
representation $R$ with a real structure; in other words,
$[R]\otimes_R\CC \cong R$.  In this case $(\repre{8},\repre{2\eN})$
has a real structure by virtue of both $\repre{8}$ and $\repre{2\eN}$
being quaternionic representations of $\so_{6,2}$ and $\sp_\eN$,
respectively.  We recognise the $\so_{6,2}$ factor as the isometry
algebra of anti-de~Sitter space.  Now we discuss the other factor.

\subsubsection{$\eN{=}2$}

In this case, the $4$-manifold is the round sphere $S^4$.  The Killing
spinors transform as spinors of $\so_5\cong\sp_2$, in agreement with
the structure of the superconformal algebra.

\subsubsection{$\eN{=}1$}

This case is obtained by considering orbifolds of the previous case.
Hence $X_4 \cong S^4/\Gamma$ where $\Gamma$ is an ADE subgroup
of the `hyperk\"ahler' $\Sp_1\subset\SO_4\subset\SO_5$.  Since the
centraliser of its Lie algebra $\sp_1$ in $\so_5$ is precisely the
other $\sp_1$ in the decomposition $\so_4 \cong \sp_1\times\sp_1$,
the resulting theory has an $\sp_1$ factor in its bosonic subalgebra
just as expected.  To see how this acts on the spinors, we must
decompose the spinor representation $\repre{4}$ of $\so_5 \cong
\sp_2$ in terms of its maximal $\sp_1\times\sp_1$ subalgebra.  We find
\begin{equation*}
\repre{4} \to (\repre{2},\repre{1}) \oplus  (\repre{1},\repre{2})~;
\end{equation*}
whence the $\Gamma$-invariant spinors transform according to the
$\repre{2}$ of $\sp_1$, in agreement with the structure of the
superconformal algebra.  (Notice that the $\repre{2}$ of $\sp_1$ is
indeed quaternionic.)

\section{Conclusions}

We have discussed spaces of the form $\ads_{p+2} \times X_{d}$ and
given a classification of the geometries of $X_d$ that admit any given
number of Killing spinors.  The geometries admitting Killing spinors
are, up to quotients: spherical, Einstein-Sasaki in dimension
$2k{+}1$, $3$-Sasaki in dimension $4k{+}3$, $7$-dimensional manifolds
of weak $G_2$ holonomy and $6$-dimensional nearly K\"ahler manifolds.
When the space $\ads_{p+2} \times X_{d}$ is a solution of a
supergravity theory, our analysis gives the amount of supersymmetry
the solution preserves.  Our approach was based on considering the
corresponding brane solution which interpolates between an $\ads_{p+2}
\times X_{d}$ near-horizon geometry and the solution $M_{p+1}\times
C(X)$ which is a product of $p+1$-dimensional Minkowski space and the
cone $C(X)$ over $X$.  The Killing spinors on $X$ are related to
parallel spinors on $C(X)$, and the number of these depends only on
the holonomy of $C(X)$.  In this way, we are able to relate the number
of supersymmetries of the two asymptotic regions to each other and to
the number of supersymmetries of the brane solution that interpolates
between them.  We have found the anti-de Sitter supergroup of
symmetries that emerges for each supergravity solution.

The brane picture leads to a conjectured duality between the string or
$\M$ theory on $\ads_{p+2} \times X_{d}$ and a superconformal field
theory with the same supergroup, which we propose arises as the
world-volume theory of $p+1$ branes located at the conical singularity
of $M_{p+1}\times C(X)$.  The geometry then gives rise to some
interesting predictions for the properties of the superconformal field
theory duals.  We finish with some speculations concerning these dual
field theories.

In \cite{KlebanovWitten}, the IIB theory on $\ads _5\times X_5$ was
considered, where $X_5$ is a coset space whose cone $C(X_5)$ is a
conically singular Calabi--Yau threefold, and a dual super Yang--Mills
theory was proposed, representing $\D3$-branes at the conical
singularity.  Since this Calabi--Yau space has a description as a
K\"ahler quotient, the quantum field theory was defined as one whose
moduli space {\em is\/} this quotient. In other words the Calabi--Yau
space is identified as the Higgs branch of the Yang--Mills theory. In
particular, since the Calabi--Yau space is K\"ahler, the Yang--Mills
theory must have four-dimensional $\eN{=}1$ supersymmetry.

Many of the conical geometries we have considered also arise from
quotient constructions, and this suggests a similar structure in the
field theory duals.  For example, for the $\M2$-branes, the cones
$C(X)$ are eight-dimensional spaces with $\Sp_2$, $\SU_4$ or $\Spin_7$
holonomy. Of the examples discussed in section five, {\em all\/} the
cones of $\Sp_2$ holonomy have a description as hyperk\"ahler
quotients \cite{HKLR} and at least some (perhaps all) of those with
$\SU_4$ holonomy have descriptions as K\"ahler quotients.  Similarly,
most of the cones with $\SU_3$ holonomy are K\"ahler quotients.

We propose, in the cases in which $C(X)$ is obtained via a quotient
construction in which a space $M$ is `divided´ by a group $G$, that
the corresponding superconformal field theories have scalars taking
values in $M$, and that there is a scalar potential whose space of
minima, i.e., the moduli space of the theory, is precisely the cone
$C(X)$.  It is natural to identify $G$ as the gauge group of the
theory.  In the (hyper-)K\"ahler quotient, $M$ is essentially the zero
level set of the moment map defined by $G$.  This generalises the
construction of \cite{KlebanovWitten}.

For example, consider the 3-dimensional case.  Recall that the
$\M2$-brane theory arises as an infrared fixed point of the $d{=}3$
super Yang--Mills theory that is the world-volume theory of the
$\D2$-brane.  Consider those (non-maximally supersymmetric)
3-dimensional superconformal field theories that arise as infrared
fixed points of $d{=}3$ super Yang--Mills theories, and which are also
the world-volume theories for $\M2$-branes on cones with $\Sp_2$ or
$\SU_4$ holonomy.  The quotients are hyperk\"ahler or K\"ahler, and
the $d{=}3$ Yang--Mills theories will have $\eN{=}4$ or $\eN{=}2$
supersymmetry, respectively.  We can choose a potential whose zeroes
define a moduli space which is precisely the cone $C(X)$, and a
natural proposal is that these Yang--Mills theories flow in the
infrared to superconformal field theories which are holographically
dual to $\M$ theory on $\ads_4\times X_7$.

The geometries we have found point to a number of situations in which
the superconformal field theory dual would be particularly interesting
to understand. For the $\M2$-branes with $X_7$ 3-Sasakian, the dual
theory should be some theory with $\eN{=}3$ superconformal symmetry,
which would be interesting to construct directly.

We saw in section 5 that any $3$-Sasakian $7$-manifold may be squashed
metrically to give a second Einstein metric with weak $G_2$ holonomy.
Let $g$ be the $3$-Sasakian metric and $\Tilde g$ the associated
squashed metric on the $7$-manifold $X_7$. Then $\M$ theory
compactified on $(X_7,g)$ has $\eN{=}3$ supersymmetry in four
dimensions (or $\eN{=}8$ supersymmetry for the round sphere), while it
has $\eN{=}1$ when compactified on $(X_7,\Tilde g)$.  When $g$ is the
round metric on the $7$-sphere, the phenomenon of squashing was
interpreted in the Kaluza-Klein literature \cite{DNP} as spontaneous
(super)symmetry breaking.  One can of course give a similar
interpretation for any $X_7$ which admits a $3$-Sasakian metric.  From
the two $\M$ theory compactifications specified by $g$ and $\Tilde g$
the adS/CFT correspondence gives rise to two corresponding
superconformal field theories in three dimensions. It is natural to
expect a relation between this pair of SCFTs. The squashing of the
$3$-Sasakian metric to the weak $G_2$ holonomy metric is a continuous
deformation, but the resulting one-parameter family of metrics is
Einstein only at two points. Of this one-dimensional family of
metrics, only two are solutions to $\M$ theory.  This suggests a
corresponding one-parameter family of supersymmetric field theories
which is superconformal at only two special values of the parameter,
with $\eN{=}3$ (or $\eN{=}8$) superconformal symmetry at one point and
$\eN{=}1$ superconformal symmetry at the other.  Understanding the
precise relation between this pair of superconformal field theories
should also prove interesting mathematically, since the moduli space
of the $\eN{=}1$ theory apparently has $\Spin_7$ holonomy.

\section*{Acknowledgements}
It is a pleasure to thank Krzysztof Galicki, Alexandros Kehagias, Igor
Klebanov and Edward Witten for useful correspondence, and Mohab Abou
Zeid for a careful reading of the manuscript.  The work of BSA is
funded by a PPARC Postdoctoral Fellowship, that of JMF by an EPSRC
PDRA, that of CMH by an EPSRC Senior Fellowship and that of BS by an
EPSRC Advanced Fellowship.  We would like to take this opportunity to
thank the respective research councils for their continued support.

\bibliographystyle{amsplain}
\bibliography{Sugra,CaliGeo,ESYM,AdS}

\providecommand{\bysame}{\leavevmode\hbox to3em{\hrulefill}\thinspace}
\begin{thebibliography}{10}

\bibitem{AOY-MCFT}
O~Aharony, Y~Oz, and Z~Yin, \emph{M theory on {$AdS_p\times S^{11-p}$} and
  superconformal field theories}, {\tt hep-th/9803051}.

\bibitem{AOT-orbi}
C~Ahn, K~Oh, and R~Tatar, \emph{Orbifolds of {$AdS_7\times S^4$} and six
  dimensional $(0,1)$ {SCFT}}, {\tt hep-th/9804093}.

\bibitem{AloffWallach}
S~Aloff and NR~Wallach, \emph{An infinite family of distinct 7-manifolds
  admitting positively curved riemannian structures}, Bull. Am. Math. Soc.
  \textbf{81} (1975), 93--97.

\bibitem{Baer}
C~B{\"a}r, \emph{Real {K}illing spinors and holonomy}, Comm. Math. Phys.
  \textbf{154} (1993), 509--521.

\bibitem{M9brane}
E~Bergshoeff and JP~van~der Schaar, \emph{On {$\M9$}-branes}, {\tt
  hep-th/9806069}.

\bibitem{Berkooz-S4orbi}
M~Berkooz, \emph{A supergravity dual of a $(1,0)$ field theory in six
  dimensions}, {\tt hep-th/9802195}.

\bibitem{Besse}
AL~Besse, \emph{Einstein manifolds}, Springer-Verlag, 1987.

\bibitem{BD}
R~Bielawsi and AS~Dancer, \emph{The geometry and topology of toric
  hyperk\"ahler manifolds}, MPI and McMaster preprint, 1996.

\bibitem{Bielawski}
R~Bielawski, \emph{Betti numbers of 3-{S}asakian quotients of spheres by tori},
  MPI preprint, 1997.

\bibitem{BoGa}
CP~Boyer and K~Galicki, \emph{3-{Sasakian} {M}anifolds}, preprint, 1998.

\bibitem{BGM}
CP~Boyer, K~Galicki, and BM~Mann, \emph{Quaternionic reduction and {E}instein
  manifolds}, Comm. Anal. Geom. \textbf{1} (1993), 1--51.

\bibitem{BGM3SQ}
\bysame, \emph{The geometry and topology of 3-{S}asakian manifolds}, J. reine
  angew. Math. \textbf{455} (1994), 183--220.

\bibitem{BGMR}
CP~Boyer, K~Galicki, BM~Mann, and E~Rees, \emph{Compact 3-{S}asakian
  7-manifolds with arbitrary second {B}etti number}, Invent. math. \textbf{131}
  (1998), 321--344.

\bibitem{CCDAFFT}
L~Castellani, A~Ceresole, R~D'Auria, S~Ferrara, P~Fr\'e, and M~Trigiante,
  \emph{{$G/H$} {$\M$}-branes and $\ads_{p+2}$ geometries}, {\tt
  hep-th/9803039}.

\bibitem{CDAF}
L~Castellani, R~D'Auria, and P~Fr\'e, \emph{{$\SU(3)\times\SU(2)\times\U(1)$}
  from $d{=}11$ supergravity}, Nucl. Phys. \textbf{B239} (1984), 610--652.

\bibitem{CastellaniRomans}
L~Castellani and LJ~Romans, \emph{{$N{=}3$} and {$N{=}1$} supersymmetry in a
  new class of solutions for $d{=}11$ supergravity}, Nucl. Phys. \textbf{B238}
  (1984), 683.

\bibitem{CastellaniRomansWarner}
L~Castellani, LJ~Romans, and NP~Warner, \emph{A classification of compactifying
  solutions for $d{=}11$ supergravity}, Nucl. Phys. \textbf{B241} (1984),
  429--462.

\bibitem{CheegerEbin}
J~Cheeger and D~Ebin, \emph{Comparison theorems in {R}iemannian {G}eometry},
  North-Holland, Amsterdam, 1975.

\bibitem{CJS}
E~Cremmer, B~Julia, and J~Scherk, \emph{Supergravity in eleven dimensions},
  Phys. Lett. \textbf{76B} (1978), 409--412.

\bibitem{DAFvN}
R~D'Auria, P~Fr\'e, and P~van Nieuwenhuizen, \emph{{$N{=}2$} matter coupled to
  supergravity from compactifications on a coset {$G/H$} possessing an
  additional {K}illing vector}, Phys. Lett. \textbf{136B} (1984), 347--353.

\bibitem{DuffGibbonsTownsend}
MJ~Duff, GW~Gibbons, and PK~Townsend, \emph{Macroscopic superstrings as
  interpolating solitons}, Phys. Lett. \textbf{332B} (1994), 321--328, {\tt
  hep-th/9405124}.

\bibitem{DuffUntwisted}
MJ~Duff, H~L\"u, and CN~Pope, \emph{{$\mathrm{AdS}_5 \times S^5$} untwisted},
  {\tt hep-th/9803061}.

\bibitem{DLPS}
MJ~Duff, H~L\"u, CN~Pope, and E~Sezgin, \emph{Supermembranes with fewer
  supersymmetries}, Phys. Lett. \textbf{371B} (1996), 206--214, {\tt
  hep-th/9511162}.

\bibitem{DNP}
MJ~Duff, BEW Nilsson, and CN~Pope, \emph{Kaluza--{K}lein supergravity}, Phys.
  Rep. \textbf{130} (1986), 1--142.

\bibitem{DS2brane}
MJ~Duff and KS~Stelle, \emph{Multi-membrane solutions of {$D{=}11$}
  supergravity}, Phys. Lett. \textbf{253B} (1991), 113--118.

\bibitem{Gomis2}
R~Entin and J~Gomis, \emph{Spectrum of chiral operators in strongly coupled
  gauge theories}, {\tt hep-th/9804060}.

\bibitem{JMFSUSY98}
JM~Figueroa-O'Farrill, \emph{Near-horizon geometries of supersymmetric branes},
  {\tt hep-th/9807149}.

\bibitem{F4}
T~Friedrich, \emph{A remark on the first eigenvalue of the {D}irac operator on
  4-dimensional manifolds}, Math. Nach. \textbf{102} (1981), 53--56.

\bibitem{FK5}
T~Friedrich and I~Kath, \emph{Eintein manifolds of dimension five with small
  eigenvalue of the {D}irac operator}, J. Diff. Geom. \textbf{29} (1989),
  263--279.

\bibitem{FK7-2}
\bysame, \emph{7-dimensional compact riemannian manifolds with {K}illing
  spinors}, Comm. Math. Phys. \textbf{133} (1990), 543--561.

\bibitem{FKMS}
T~Friedrich, I~Kath, A~Moroianu, and U~Semmelmann, \emph{On nearly parallel
  {$G_2$}-structures}, J. Geom. Phys. \textbf{23} (1997), 259--286.

\bibitem{GalickiSalamon}
K~Galicki and S~Salamon, \emph{On {B}etti numbers of 3--{S}asakian manifolds},
  Geom. Ded. \textbf{63} (1996), 45--68.

\bibitem{Gallot}
S~Gallot, \emph{Equations diff\'erentielles caract\'eristiques de la sph\`ere},
  Ann. Sci. \'Ecole Norm. Sup. \textbf{12} (1979), 235--267.

\bibitem{GGPT}
JP~Gauntlett, GW~Gibbons, G~Papadopoulos, and PK~Townsend,
  \emph{Hyper-{K\"ahler} manifolds and multiply intersecting branes}, Nucl.
  Phys. \textbf{B500} (1997), 133--162, {\tt hep-th/9702202}.

\bibitem{GibbonsTownsend}
GW~Gibbons and PK~Townsend, \emph{Vacuum interpolation in supergravity via
  $p$-branes}, Phys. Rev. Lett. \textbf{71} (1993), 3754--3757, {\tt
  hep-th/9302049}.

\bibitem{Gomis1}
J~Gomis, \emph{Anti-de~{S}itter geometry and strongly coupled gauge theories},
  {\tt hep-th/9803119}.

\bibitem{Gray-KSAH}
A~Gray, \emph{K\"ahler submanifolds of homogeneous almost hermitian manifolds},
  T\^ohoku Math. J \textbf{21} (1969), 190--194.

\bibitem{GrayNK}
\bysame, \emph{Nearly {K}\"ahler manifolds}, J. Diff. Geom. \textbf{4} (1970),
  283--309.

\bibitem{Gray-WH}
\bysame, \emph{Weak holonomy groups}, Math. Z. \textbf{123} (1971), 290--300.

\bibitem{Gray-3S}
\bysame, \emph{Riemannian manifolds with geodesic symmetries of order 3}, J.
  Diff. Geom. \textbf{7} (1972), 343--369.

\bibitem{GrayNK2}
\bysame, \emph{The structure of nearly {K}\"ahler manifolds}, Math. Ann.
  \textbf{223} (1976), 233--248.

\bibitem{AlmostHermitian}
A~Gray and LM~Hervella, \emph{The sixteen classes of almost {H}ermitian
  manifolds and their linear invariants}, Annali di Mat. Pura Appl. (IV)
  \textbf{123} (1980), 35--58.

\bibitem{KKmonopoleGP}
DJ~Gross and MJ~Perry, \emph{Magnetic monopoles in {K}aluza--{K}lein theories},
  Nucl. Phys. \textbf{B226} (1983), 29.

\bibitem{Grunewald}
R~Grunewald, \emph{Six-dimensional manifolds with a real {K}illing spinor},
  Ann. Global Anal. Geom. \textbf{8} (1990), 43--59.

\bibitem{Gubser-Einstein}
SS~Gubser, \emph{Einstein manifolds and conformal field theories}, {\tt
  hep-th/9807164}.

\bibitem{GKP}
SS~Gubser, IR~Klebanov, and AM~Polyakov, \emph{Gauge theory correlators from
  noncritical string theory}, Phys. Lett. \textbf{428B} (1998), 105, {\tt
  hep-th/9802109}.

\bibitem{Guven}
R~G\"uven, \emph{Black $p$-brane solutions of {$D{=}11$} supergravity theory},
  Phys. Lett. \textbf{276B} (1992), 49--55.

\bibitem{HLS}
R~Haag, JT~{\L}opusza\'nski, and M~Sohnius, \emph{All possible generators of
  supersymmetries of the {$S$}-matrix}, Nucl. Phys. \textbf{B88} (1975),
  257--274.

\bibitem{HKLR}
NJ~Hitchin, A~Karlhede, U~Lindstr{\"o}m, and M~Ro{\v c}ek,
  \emph{Hyperk{\"a}hler metrics and supersymmetry}, Comm. Math. Phys.
  \textbf{108} (1987), 535--589.

\bibitem{HorowitzStrominger}
GT~Horowitz and A~Strominger, \emph{Black strings and $p$-branes}, Nucl. Phys.
  \textbf{B360} (1991), 197--209.

\bibitem{Mwave}
CM~Hull, \emph{Exact $pp$ wave solutions of eleven-dimensional supergravity},
  Phys. Lett. \textbf{139B} (1984), 39.

\bibitem{Gbranes}
\bysame, \emph{Gravitational duality, branes and charges}, Nucl. Phys.
  \textbf{B509} (1998), 216--251, {\tt hep-th/9705162}.

\bibitem{IMSY}
N~Itzhaki, JM~Maldacena, J~Sonnenschein, and S~Yankielowicz, \emph{Supergravity
  and the large {$N$} limit of theories with sixteen supercharges}, {\tt
  hep-th/9802042}.

\bibitem{KaSi-orbifolds}
S~Kachru and E~Silverstein, \emph{4d conformal field theories and strings on
  orbifolds}, {\tt hep-th/9802183}.

\bibitem{Kehagias}
A~Kehagias, \emph{New type {IIB} vacua and their {F}-theory interpretation},
  {\tt hep-th/9805131}.

\bibitem{KlebanovWitten}
IR~Klebanov and E~Witten, \emph{Superconformal field theory on threebranes at a
  {C}alabi--{Y}au singularity}, {\tt hep-th/9807080}.

\bibitem{KreckStolz}
M~Kreck and S~Stolz, \emph{A diffeomorphism classification of 7-dimensional
  homogeneous {E}instein manifolds with {$\SU(3)\times \SU(2) \times
  \U(1)$}-symmetry}, Ann. of Math. \textbf{127} (1988), 373--388.

\bibitem{LNV}
A~Lawrence, N~Nekrasov, and C~Vafa, \emph{On conformal field theories in four
  dimensions}, {\tt hep-th/9803015}.

\bibitem{Malda}
JM~Maldacena, \emph{The large {$N$} limit of superconformal field theories and
  supergravity}, {\tt hep-th/9711200}.

\bibitem{Moroianu}
A~Moroianu, \emph{On the infinitesimal isometries of manifolds with {K}illing
  spinors}, Humboldt Universit\"at preprint.

\bibitem{Morrison}
DR~Morrison, \emph{Non-spherical horizons}, Talk at Strings '98.

\bibitem{Nahm}
W~Nahm, \emph{Supersymmetries and their representations}, Nucl. Phys.
  \textbf{B135} (1978), 149--166.

\bibitem{OT-orbi}
Y~Oz and J~Terning, \emph{Orbifolds of {$AdS_5\times S^5$} and $4d$ conformal
  field theories}, {\tt hep-th/9803167}.

\bibitem{PopeWarner}
CN~Pope and NP~Warner, \emph{Two new classes of compactifications of $d{=}11$
  supergravity}, Class. \& Quantum Grav. \textbf{2} (1985), L1--L6.

\bibitem{Romans5}
LJ~Romans, \emph{New compactifications of chiral {$N{=}2$}, $d{=}10$
  supergravity}, Phys. Lett. \textbf{153B} (1985), 392.

\bibitem{Salamon}
S~Salamon, \emph{Riemannian geometry and holonomy groups}, Pitman Research
  Notes in Mathematics, no. 201, Longman Scientific \& Technical, 1989.

\bibitem{Sasaki}
S~Sasaki, \emph{On differentiable manifolds with certain structures which are
  closely related to almost contact structures}, T\^ohoku Math. J \textbf{2}
  (1960), 459--476.

\bibitem{KKmonopoleS}
RD~Sorkin, \emph{Kaluza-{K}lein monopole}, Phys. Rev. Lett. \textbf{51} (1983),
  87--90.

\bibitem{S5/G}
S~Sulanke, \emph{Der erste {E}igenwert des {D}irac-{O}perators auf
  {$S^5/\Gamma$}}, Math. Nach. \textbf{99} (1980), 259--271.

\bibitem{Wang}
MY~Wang, \emph{Parallel spinors and parallel forms}, Ann. Global Anal. Geom.
  \textbf{7} (1989), no.~1, 59--68.

\bibitem{WittenAdS}
E~Witten, \emph{Anti de sitter space and holography}, {\tt hep-th/9802150}.

\bibitem{WittenBaryons}
\bysame, \emph{Baryons and branes in anti-de~{S}itter space}, {\tt
  hep-th/9805112}.

\bibitem{WittenMpqr}
\bysame, \emph{Search for a realistic {K}aluza--{K}lein theory}, Nucl. Phys.
  \textbf{B186} (1981), 412--428.

\bibitem{Wolf}
JA~Wolf, \emph{Spaces of constant curvature}, {T}hird ed., Publish or Perish,
  Boston, 1974.

\end{thebibliography}
\end{document}